\documentclass[fleqn]{mnras}

\usepackage{newtxtext,newtxmath}

\usepackage[T1]{fontenc}
\DeclareUnicodeCharacter{2212}{-}

\DeclareRobustCommand{\VAN}[3]{#2}
\let\VANthebibliography\thebibliography
\def\thebibliography{\DeclareRobustCommand{\VAN}[3]{##3}\VANthebibliography}

\usepackage{graphicx}	

\usepackage{amsmath}	
\usepackage{amssymb}	

\title[Decoding NGC 628]
{Decoding NGC~628 with radiative transfer methods}

\author[M. T. Rushton et al.]{
M. T. Rushton$^{1,2,3}$\thanks{E-mail: mrushton@uclan.ac.uk},
C. C. Popescu$^{2}$,
C. Inman$^{2}$,
G. Natale$^{2}$,
and D. Pricopi$^{1}$
\\
$^{1}$Astronomical Institute of the Romanian Academy, Str. Cutitul de Argint 5, 040557, Bucharest, Romania\\
$^{2}$Jeremiah Horrocks Institute, University of Central Lancashire, Preston, PR1 2HE, UK \\
$^{3}$Max Planck Institute f\"{u}r Kernphysik, Saupfercheckweg 1, D-69117 Heidelberg, Germany
}

\date{Accepted XXX. Received YYY; in original form ZZZ}

\pubyear{2015}

\begin{document}
\label{firstpage}
\pagerange{\pageref{firstpage}--\pageref{lastpage}}
\maketitle

\begin{abstract}
We present an axi-symmetric  model for the ultraviolet (UV)-to-submillimetre (submm) images of the nearly face-on spiral galaxy NGC~628. It was calculated using a radiative transfer (RT) code, accounting for the absorption and re-emission of starlight by dust in the interstellar medium of this galaxy. The code incorporates emission from Polycyclic Aromatic Hydrocarbons, anisotropic scattering and stochastic heating of the grains. This is the second successful modelling of a face-on spiral galaxy with RT  methods, whereby the large-scale geometry of stars and dust is self-consistently determined. The solution was obtained by fitting azimuthally averaged profiles in the UV, optical and submm. The model predicts remarkably well all characteristics of the profiles, including the increase by a factor of 1.8 of the scale-length of the infrared emissivity between 70 and 500\,${\mu}$m. We find that NGC~628 did not undergo an efficient inside-out disk growth, as predicted by semi-analytical hierarchical models for galaxy formation. We also find large amounts of dust grains at large radii, that could involve efficient transport mechanisms from the inner disk.
Our results show that $71\%$ of the dust emission in NGC~628 is powered by the young stellar populations, with the old stellar populations from the bulge contributing $65\%$ to the heating of the dust in the central region ($R<0.5$\,kpc). 
The derived star formation rate is $\rm SFR=2.00\pm0.15\,{\rm M}_{\odot}{\rm yr}^{-1}$.  

\end{abstract}




\section{Introduction}

Rotationally-supported disk galaxies are the fundamental building blocks of the visible Universe. They, and their high-{\it z} predecessors, host the bulk of star formation (SF) over the history of the Universe. Disks act as areas where gas accretes from the surrounding Intergalactic Medium (IGM) (e.g. Sanchez Almeida et al. 2014, Putman 2017, Ho et al. 2019), but the very gas out of which stars are forming is dusty, and as such it  prevents a clear view of the SF process. This, in turn, prevents a straightforward derivation of  the SF history of disk galaxies. To derive the intrinsic properties of galaxies, a quantitative understanding of the dust effects is needed. 

Dust absorbs and scatters the stellar light, with efficiencies that depend on the composition, shape and size of the grains. These efficiencies can be calculated using the Mie theory, since the size of the grains is usually comparable with the wavelength of the stellar light. The absorption and scattering efficiencies decrease with increasing wavelength, although not always monotonic. The combined effect of absorption and scattering  - what we call extinction - thus fully characterizes the interaction of dust grains with radiation. When looking at an extended distribution of emitters (stars) and absorbers (dust), as is the case of a galaxy, the overall absorption and scattering of the light will not only depend on the extinction properties of the dust, but also on the spatial distribution of stars and dust. Thus, on any line of sight, light can not only be diminished by absorption and scattering away from the line of sight, but also amplified by scattered light from other directions into the line of sight. This effect is called attenuation.

Dust re-emits the absorbed direct stellar light as thermal radiation in the infrared, with a spectral energy distribution that depends on the emission efficiencies of the grains and their temperature. While some of the grains are heated at equilibrium temperature by their ambient radiation fields, a large majority of dust grains would never reach a thermal balance, but instead exhibit  temperature fluctuations. This latter effect occurs when the cooling timescale of the dust is smaller than the timescale for consecutive photons hits, and arises  if the grain is small and the energy density of the radiation fields is low. The grains that exhibit temperature fluctuations are said to be stochastically heated.

The complexity of the processes related to dust attenuation and 
emission in spiral galaxies means that there is no simple recipe  to account for them. In particular the effects of dust attenuation are severe, especially in the ultraviolet (UV) range, 
or in regions of high opacity, such as the inner regions of disks or spiral arms, not only in the UV, but even in the optical (Popescu 2021). Dust attenuation can have a significant impact on face-on galaxies as well. Examples include the change in the apparent UV surface brightness distributions with respect to what it would be observed in the absence of dust (see Fig. 15 from Thirlwall et al. 2020). It was also shown that the variation of attenuation with wavelength (the attenuation curve) derived at face-on orientations changes with radial position (Thirwall et al. 2020), simply due to the radial gradient in dust opacity.

Being mindful that dust distorts our view of galaxies, how do we derive the intrinsic properties of galaxies? To give a precise, quantitative  answer to this question we need self-consistent models, which perform radiative transfer (RT) calculations, taking into account, and determining, the relative geometries of stars and dust. Detail RT modelling of the UV/optical/far-infrared (FIR)/submillimetre (submm) spectral energy distribution (SED) of individual spiral galaxies has been done in the past for edge-on galaxies (Popescu et al. 2000,  Misiriotis et al. 2001, Popescu et al. 2004, Bianchi 2008, Baes et al. 2010, Mosenkov et al. 2016, Popescu et al. 2017, Mosenkov et al. 2018, Natale et al. 2021), since in this orientation it is possible to derive the vertical distribution of stars and dust. More recently work has been also started to model face-on galaxies. In a face-on view there is more information on the radial distribution of stellar emissivity and dust opacity, and on the non-axisymmetric structure, but the vertical distributions have been proven difficult to constrain (Combes et al. 2012, Koch et al. 2020). Two approaches have been adopted to-date to tackle the challenge of modelling face-on galaxies. 

The first approach is to use non-axisymmetric RT models, which  present the great advantage that they can incorporate detailed 3D structure. However, their advantage comes with the disadvantage that, so far, they have only been used to fit the spatially integrated SED (as anything else would be computationally prohibited), under the assumption that the geometry of stars and dust is known and follows the observed images or is inputted from empirical energy balance models on a pixel-to-pixel implementation.   This approach has been adopted by de Looze et al. (2014) and Nersesian et al. (2020a) for M51, Viaene et al. (2017) for M31, Williams et al. (2019) for M33, Verstocken et al. (2020) for M81, Viaene et al. (2020) for NGC~1068, and Nersesian et al. (2020a) for 4 barred galaxies.

The second approach is to use axi-symmetric RT models, which have the great advantage that they can be used to fit  the geometries of stars and dust, so in this sense they provide a fully self-consistent modelling, but they have the inherent limitation imposed by the azimuthal symmetry. So far this second approach has only been adopted for modelling M33 by Thirlwall et al. (2020). While successful in fitting the available observed surface brightness profiles of M33 at all UV/optical/FIR/submm wavebands, the question arises of whether this approach can be successfully adopted for other galaxies as well. Afterall, M33 is the nearest spiral galaxy, for which extreme detail can be gained. M33 is also a bulgeless galaxy, with floculent spiral structure, which is ideal for axi-symmetric modelling. So would this approach work for a galaxy  with a different morphological structure? The goal of this paper is to address this question by providing the second self-consistent RT modelling of a face-on spiral galaxy, for the case of NGC~628. 

In addition, NGC~628 is an important target for studies spanning a broad range of topics, including studies of galaxy evolution and stellar populations (Natali et al. 1992; Cornett et al. 1994, Zou et al. 2011;  Shabani et al. 2018), star-formation (Hodge 1976; Kennicutt \& Hodge 1976; Kennicutt \& Hodge 1980; Elmegreen et al. 2006; Gusev 2014; Grasha et al. 2015; Muraoka et al. 2016; Luisi et al. 2018; Rousseau-Nepton et al. 2018; Yadav et al. 2021), supernova remnants surveys (Sonbas et al. 2010), interstellar medium (Berg et al. 2013; Crocker et al. 2013; Croxall et al. 2013;  Blanc et al. 2013; Abdullah et al. 2017; Vilchez et al. 2019) and galaxy dynamics (Honig \& Reid 2015; Inoue et al. 2020; Sakhibov et al. 2021). It is also customary to show NGC~628 as an example of how the Milky Way would appear face-on to an external observer,  simply because it is the  closest, almost face-on, grand design spiral. It has even been eluded that NGC~628 can be considered a twin of the Milky Way (Belley \& Roy 1992; Lelievre \& Roy 2000), despite the slightly different Hubble type.   Because of this NGC~628 has been incorporated in major surveys of nearby Universe galaxies: {\it Spitzer} Nearby Galaxies Survey (SINGS, Kennicut et al. 2003), the GALEX Nearby Galaxy Survey (NGS; Gil de Paz et al. 2007), {\it Spitzer} Local Volume Legacy (SLVL, Dale et al. 2009), Key Insights on Nearby Galaxies: A Far-Infrared Survey with {\it Herschel} (KINGFISH, Kennicut et al. 2011), Legacy Extragalactic UV Survey (LEGUS, Calzetti et al. 2015), The Chemical Abundances of Spirals (CHAOS, Berg et al. 2015),  Physics at High Angular Resolution in Nearby Galaxies (PHANGS, Leroy et al. 2021a,b; PHANGS-MUSE, Emsellem et al. 2022); PHANGS-HST, Lee et al. 2022). Thus, decodding the spatial and spectral energy distribution of NGC~628 is in itself an important task. 

Furthermore, the RT solution obtained for NGC~628 can be used, together with the one for M33, 
to inform simpler methods of analysing multiwavelength data, like energy balance methods or other phenomenological or empirical methods. For example, RT solutions can be used to show the effects of geometry on attenuation curves, commonly used to derive intrinsic properties.
RT solutions can be also used to calibrate SFR indicators or to quantify the dust effects on the observed surface brightnesses profiles.

NGC 628 (or Messier 074) is classified as a Sc spiral galaxy by Hubble (1926) and Sandage (1961), and as a SA(s)c by de Vaucouleurs et al. (1991). It is the dominant galaxy of a group, potentially having two dwarf galaxy satellites (Davis et al. 2021).  Otherwise NGC~628 is relatively isolated and largely unperturbed (Kamphuis et al. 1992; Auld et al. 2006). It has a faint $H\alpha$ emitting outer disk, extending beyond two optical radii (defined by the 25th B-band isophote) (Ferguson et al. 1989, Lelievre \& Roy 2000) and an extended HI disk (Briggs et al. 1980, Briggs 1982).
The distance to NGC 628 has been determined with many techniques
(Bottinelli et al. 1984; Ivanov et al. 1992; Rodriquez et al. 1994; Zasov et al 1996; Sohn et al. 1996; Sharina et al. 1996;  Vinko et al. 2004; Henry et al. 2005; Van Dyk et al. 2006; Herrmann et al. 2008;  Olivares et al. 2010; Jang et al. 2014; Kreckel et al. 2017). It is generally determined to be in the range of $7-10$\,Mpc. Recently, a distance of $9.52^{+0.26}_{-0.41}$\,Mpc  was derived by Scheuermann et al. (2022) using the planetary nebulae luminosity function distance. In this paper we adopt a distance of 9.5\,Mpc.

The structure of this paper is as follows.
In Sect.~\ref{sec:data} we present the observations used in this analysis. The structural components of the model are described in Sect.~\ref{sec:struct}. In Sect.~\ref{sec:rt} we summarise the main characteristics of the radiative transfer codes used in this paper and in Sect.~\ref{sec:optim} we describe the method we adopt to fit the model to the data. The results are presented in Sect.~\ref{sec:results}, including the model fits to the azimuthally averaged profiles, the global SED, the derived SFR, dust mass and dust opacity and the radiation fields. We discuss these results and their consequences in Sect.~\ref{sec:discussion} and we give the summary and conclusions in Sect.~\ref{sec:summary}. 

\section{Observational Data}
\label{sec:data}

\begin{center}
\begin{table*}
\caption{The NGC~628 spatially integrated observed flux densities}
  \begin{tabular}{ | l | c | c | c | c | c}
    \hline
    $\lambda_{\rm eff}$ & Instrument \& Filter/Channel &   PSF FWHM  &          $F_{\nu}^1$                &        $F_{\nu}^2$             & References$^3$\\ 
         ($\mu$m)       &                              &    (arcsecs)&               (Jy)               &          (Jy)               &       \\ 
            (1)         &            (2)               &    (3)  &          (4)                 &            (5)              &  (6) \\ \hline 
          0.15          &         GALEX FUV            &    4.2  &  $4.40\pm0.7\times10^{-2}$   &  $4.69\pm0.73\times10^{-2}$ & 1  \\ 
          0.23          &         GALEX NUV            &    4.2  &  $6.10\pm0.9\times10^{-2}$   &  $5.91\pm0.91\times10^{-2}$ & 1  \\ 
          0.35          &         SDSS u               &    2.0  &  $0.14\pm0.005$              &  $0.17\pm0.03$              & 2\\ 
          0.48          &         SDSS g               &    2.0  &  $0.41\pm0.009$              &  $0.49\pm0.09$              & 2\\ 
          0.62          &         SDSS r               &    2.0  &  $0.67\pm0.014$              &  $0.80\pm0.16$              & 2\\
          0.76          &         SDSS i               &    2.0  &  $0.91\pm0.019$              &  $1.08\pm0.02$              & 2\\ 
          0.91          &         SDSS z               &    2.0  &  $1.09\pm0.032$              &  $1.23\pm0.02$              & 2\\ 
          1.25          &         2MASS J              &    3.3  &  $1.39\pm0.039$              &  $1.56\pm0.07$              & 3\\ 
          1.66          &         2MASS H              &    3.1  &  $1.52\pm0.06$               &  $1.60\pm0.08$              & 3\\ 
          2.20          &         2MASS K              &    3,3  &  $1.26\pm0.05$               &  $1.28\pm0.06$              & 3\\ 
          3.40          &         WISE W1              &    1.7  &  $0.683\pm0.068$             &                             & 4\\ 
          3.55          &         IRAC 1               &    6.1  &  $0.731\pm0.073$             &   $0.839\pm0.117$           & 5\\ 
          4.50          &         IRAC 2               &    1.7  &  $0.497\pm0.050$             &   $0.544\pm0.75$            & 5\\ 
          4.60          &         WISE W2              &    6.4  &  $0.582\pm0.058$             &                             & 4\\ 
          5.80          &         IRAC 3               &    1.9  &  $1.416\pm0.142$            &   $1.16\pm0.14$             & 5\\ 
          8.00          &         IRAC 4               &    2.0  &   $3.577\pm0.251$            &   $2.59\pm0.33$             & 5 \\ 
          11            &         WISE W3              &    6.5  &  $9.60\pm0.96$        &                                                 & 4\\ 
          23            &         WISE W4$^4$          &    12   &  $3.22\pm0.48$        &                                                 & 4\\ 
          24            &         MIPS 24              &    6    &      $3.13\pm0.127$     &  $3.23\pm0.12$                                                 & 6\\ 
          70            &         PACS Blue            &    5.8  &      $38.56\pm6.5$     &  $42.0\pm2.1$                                                & 7 \\ 
          100           &         PACS Green           &    12.1 &      $73.36\pm9.3$     &  $83.6\pm4.1$                                                 & 7\\ 
          160           &         PACS Red             &    18.2 &      $111.9\pm11.99$    &  $114\pm5$                                                & 7\\ 
          250           &         SPIRE PSW            &    24.9 &      $64.3\pm4.517$     &  $62.1\pm4.4$                                                & 8\\ 
          350           &         SPIRE PMW            &    24.9 &      $31.07\pm2.293$     &  $29.8\pm2.1$                                           & 8\\ 
          500           &         SPIRE PLW            &    36.3 &      $12.91\pm0.97$    &  $11.4\pm0.8$                                              & 8\\ 
    \hline
\multicolumn{6}{l}{$^1$ Flux density  derived by us and corrected for galactic extinction.} \\
\multicolumn{6}{l}{$^2$ Flux density derived by Dale et al. (2017)} \\
\multicolumn{6}{l}{$^3$(1) Morrissey et al. (2007), (2) Aihara et al. (2011), (3) Skrutskie et al. (2006), (4) Wright et al. (2010), (5) Fazio et al. (2004)} \\ 
\multicolumn{6}{l}{~\,(6) Rieke et al. (2004), (7) Poglitsch et al. (2010), (8) Griffin et al. (2010)} \\
\multicolumn{6}{l}{$^4$ Omitted from the analysis because of calibration issues.}
\label{obs}
  \end{tabular}
\end{table*}
\end{center}

As a large, nearby, nearly face-on, star-forming galaxy, NGC~628 is ideal for studies on a variety of topics, as described in the previous section. It has, therefore, been thoroughly observed throughout the electromagnetic spectrum. There is plenty of publically available data for this galaxy, in particular a  high quality dataset of imaging observations spanning the UV-to-submillimetre. A summary of the observations used in this paper is given below. Included are ground-based observations from the Sloan Digital Sky Survey (SDSS) and Two Micron Sky Survey (2MASS), and  space-based observations from the Galaxy Evolution Explorer (GALEX), the Wide-field Infrared Survey Explorer (WISE), the {\it Spitzer} Space Telescope, and the {\it Herschel} Space Telescope. The dataset was downloaded from the NASA Extragalactic Database (NED)\footnote{https://ned.ipac.caltech.edu}. Examples of imaging observations of NGC~628 used in this paper can be seen in Fig.~\ref{fig:N628_Images} (left column). 

\subsection{GALEX}

NGC~628 was one of about 200 galaxies in the GALEX Nearby Galaxy Survey (NGS; Gil de Paz et al. 2007). The GALEX ultraviolet space telescope (Martin et al. 2005) made observations in the far-UV (FUV; $\lambda_{\rm eff}=0.15\,\mu$m) and near-UV (NUV; $\lambda_{\rm eff}=0.23\,\mu$m). The count rate was converted into flux density units using the FUV and NUV zero point magnitudes (Morrissey et al. 2007).

\subsection{SDSS}

Images of NGC~628 were included in Data Release 12 of SDSS. In SDSS terminology, the field of the galaxy was observed in Run 7845, Camcol 2 and Field 104. Images were obtained in the $ugriz$ bands. SDSS expresses flux in nanomaggies (nMgy). The images were converted from counts to nMgy using the calibration factor in the FITS Header, and then to units of flux density. 

\subsection{2MASS}

2MASS was a ground-based all-sky survey in the near-infrared $JHK$ bands (Jarrett et al. 2003). The angular resolution was estimated from the FWHM of the foreground star intensities in the 29$'$ field of view (Table~\ref{obs}). The data was converted from units of counts to units of flux density using the calibration factors in the FITS Headers. 

\subsection{WISE}

WISE conducted an all-sky survey in four infrared bands, at 3.4\,$\mu$m (W1), 4.6\,$\mu$m (W2), 12\,$\mu$m (W3), and 22\,$\mu$m (W4). The data uploaded to NED had been processed and made available in pixel units of WISE magnitudes. We converted the units into flux density units using the zero magnitude flux density (uncertainity $\pm1.5$\%) from Wright et al. (2010). The WISE calibration is based on bright stars and colour corrections should therefore be significant for star-forming galaxies, especially in the wide W3 filter. The W4 spectral response is uncertain for sources with a rising infrared spectrum, and a correction is recommended, up to 10\%. This filter overlaps with one obtained by {\it Spitzer}, and we omit it from our analysis because of the uncertainty in the response.

\subsection{\it Spitzer}

{\it Spitzer} observed NGC~628 as part of the {\it Spitzer} Nearby Galaxies Survey (SINGS; Kennicutt et al. 2003). It had two onboard imagers: Infrared Array Camera (IRAC; Fazio et al. 2004) and Multiband Imaging Photometer (MIPS; Rieke et al. 2004). IRAC had four channels, simultaneously imaging at 3.6\,$\mu$m, 4.5\,$\mu$m, 5.8\,$\mu$m and 8.0\,$\mu$m. MIPS imaged in three bands, centred at about 24\,$\mu$m, 70\,$\mu$m, and 160\,$\mu$m. The last two overlap with two of {\it Herschel}'s bands, and were not used in this analysis.

\subsection{\it Herschel}

KINGFISH (Key Insights on Nearby Galaxies: A Far-Infrared Survey with {\it Herschel}; Kennicutt et al. 2011) was a panochromatic survey of 61 SINGS galaxies, including NGC~628. Images were obtained with {\it Herschel}'s Photoconductor Array Camera and Spectrometer (PACS; Poglitsch et al. 2010) and Spectral and Photometric Imaging Receiver (SPIRE; Griffin et al. 2010). PACS observed near the peak of the thermal dust emission in star-forming galaxies (70\,$\mu$m, 100\,$\mu$m and 160\,$\mu$m), and SPIRE observed the Rayleigh-Jeans tail (250\,$\mu$m, 350\,$\mu$m and 500\,$\mu$m). The {\it Herschel} mirror was about four times larger than that of {\it Spitzer}. The spatial resolution is, therefore, four times superior at 70\,$\mu$m and 160\,$\mu$m.

\subsection{Galactic extinction}

NGC628 is well south of the galactic plane on the sky ($b=-45.7$) and viewed through low galactic extinction. The UV-to-near IR data were dereddened using $E(B-V)=0.06$
(Schlafy et al. 2011). We assume the galactic extinction curve of Cardelli et al. (1989) and total-to-selective extinction ratio $R_V=3.1$. 

\subsection{Surface photometry}

Foreground stars were identified and masked using a routine for creating median-filtered images and calculating contrast. To quality as a star, a candidate had to be brighter than the local brightness multiplied by a user-defined factor. The masked areas were circular, centred on the stars, and with radii depending on the stars brightnesses. The stars were divided into two groups: bright stars requiring large masked areas and faint stars requiring small masked areas. The brightness level delimiting the grouping, and the masked area radii, were judged by eye. Further masking for bad pixels was necessary for some data. No notable diffraction spikes were present in the images. Stars were masked up to the MIR range.

We defined a set of nested ellipses (semi-minor to semi-major axis ratio $b/a=0.94$, as measured by us for NGC~628) centred on the galaxy and averaged the emission within each annulus to produce an azimuthally averaged observed image and a surface brightness profile. Spatially integrated flux (spectral) densities $F_{\nu}$ were obtained from the curves of growth (CoG) - the cumulative total of the annuli defined when constructing the surface brightness profiles. These are given in column 4 of Table~\ref{obs}. Fluxes of NGC~628 determined by Dale et al. (2017) are shown in column 5 of Table~\ref{obs} for comparison purposes only. Their values are consistent with many of those determined here, but there are some exceptions. The Dale et al. fluxes were derived using aperture photometry while we used CoG analysis. Discrepancies in the estimated sky values and differences in the masking could give inconsistent results. In all cases we used our own derived  photometry.

The errors in the spatially integrated flux densities are calculated in the same manner as Thirlwall et al. (2020) (see their Appendix A) and are listed in column 4 of Table~\ref{obs}, together with the corresponding derived flux densities. We combine in quadrature the calibration error, the error due to background fluctuations and the Poisson noise to obtain the total errors in the integrated flux densities.\\

\noindent
{\it Background fluctuations}\\
Most of the images are centred (or nearly centred) on NGC~628 and are sufficiently large for full coverage of the 2$\pi$ of azimuth angle at radii dominated by background noise. For this data, the background noise is readily measured from the surface brightness profiles. Uncertainties are relative large, for example in the IRAC images, on which the noise clearly varies across the array. As in Thirlwall et al. (2020), we determine a mean background level, derived from the total counts in 6 annuli, just beyond the visible extent of the galaxy. The background fluctuations were calculated from the annulus-to-annulus variation.\\ 
\begin{figure*}
\includegraphics[width=0.7\textwidth,angle=-0]{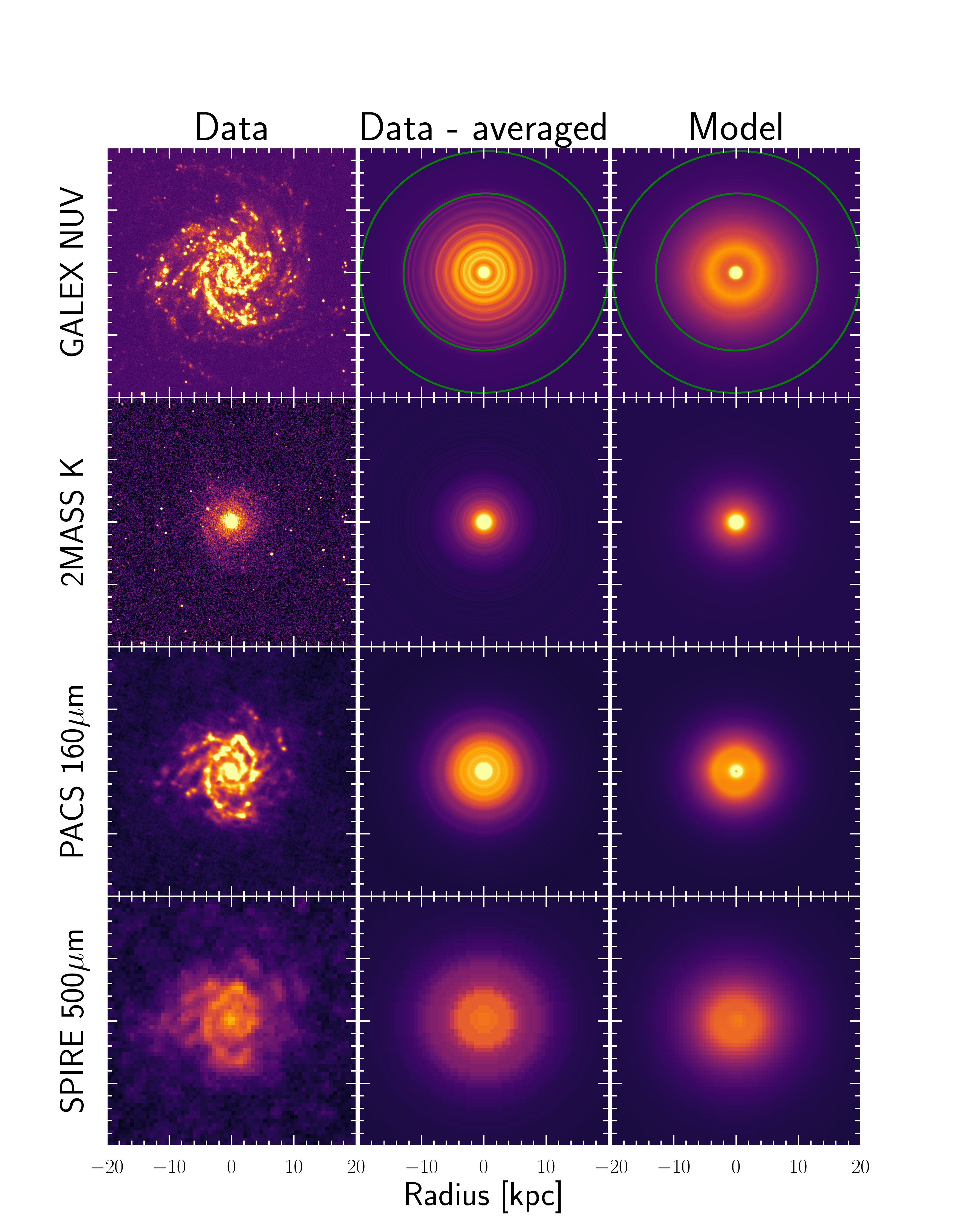}
\caption{ Observed (left), azimuthally averaged observed (middle) and model (right) images of NGC~628 at selected wavebands. For the NUV observed averaged and model images we indicate with two green elipses, placed at 13 and 20\,kpc, respectively, the position of a faint outer disk traced in the radial profiles, but less visible in the image display.}
\label{fig:N628_Images}
\end{figure*}

\noindent
{\it Poisson noise}\\
The Poisson noise was calculated in the usual way, using $\sqrt{C}$, where C is the total number of counts from  the galaxy, determined from the background-subtracted CoG. The Poisson noise is in general a small source of error, and is negliglible at longer wavelengths.\\ 

\begin{figure*}
  \centering
  \includegraphics[width=0.8\linewidth]{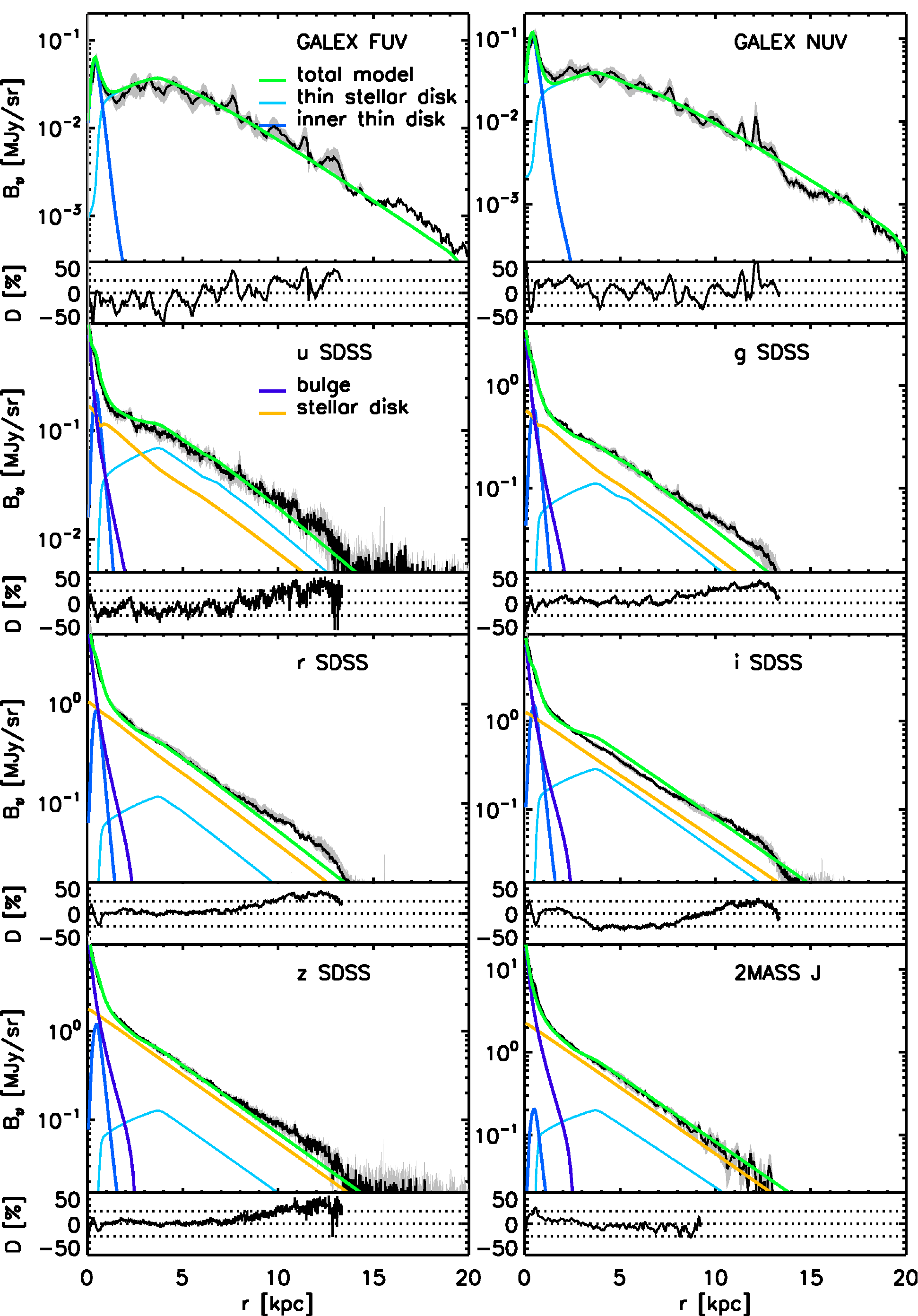}
\caption{UV/optical azimuthally average surface brightness radial profiles of NGC~628 (black line), with the associated errors shown as grey shades. The best-fitting model profiles are also shown superimposed. The total model emission is shown by the green line. Its components are also shown: the thin stellar disk (light blue), the inner thin stellar disk (dark blue), the stellar bulge (purple) and the stellar disk (orange). Bottom panels: The residuals (D), defined as the percentage difference between the data and the model, $D=(B_{\rm obs}-B_{\rm mod})$\ /$B_{\rm obs}\times100$, where $B_{\rm obs}$ is the observed average brightness in each annulus and $B_{\rm mod}$ is the model average brightness in each annulus. Dashed-lines are shown at $\pm20$\% to guide the eye.}
 \label{fig:UVopt_sbp}
\end{figure*}
 
  \begin{figure*}
  \centering
  \includegraphics[width=0.8\linewidth]{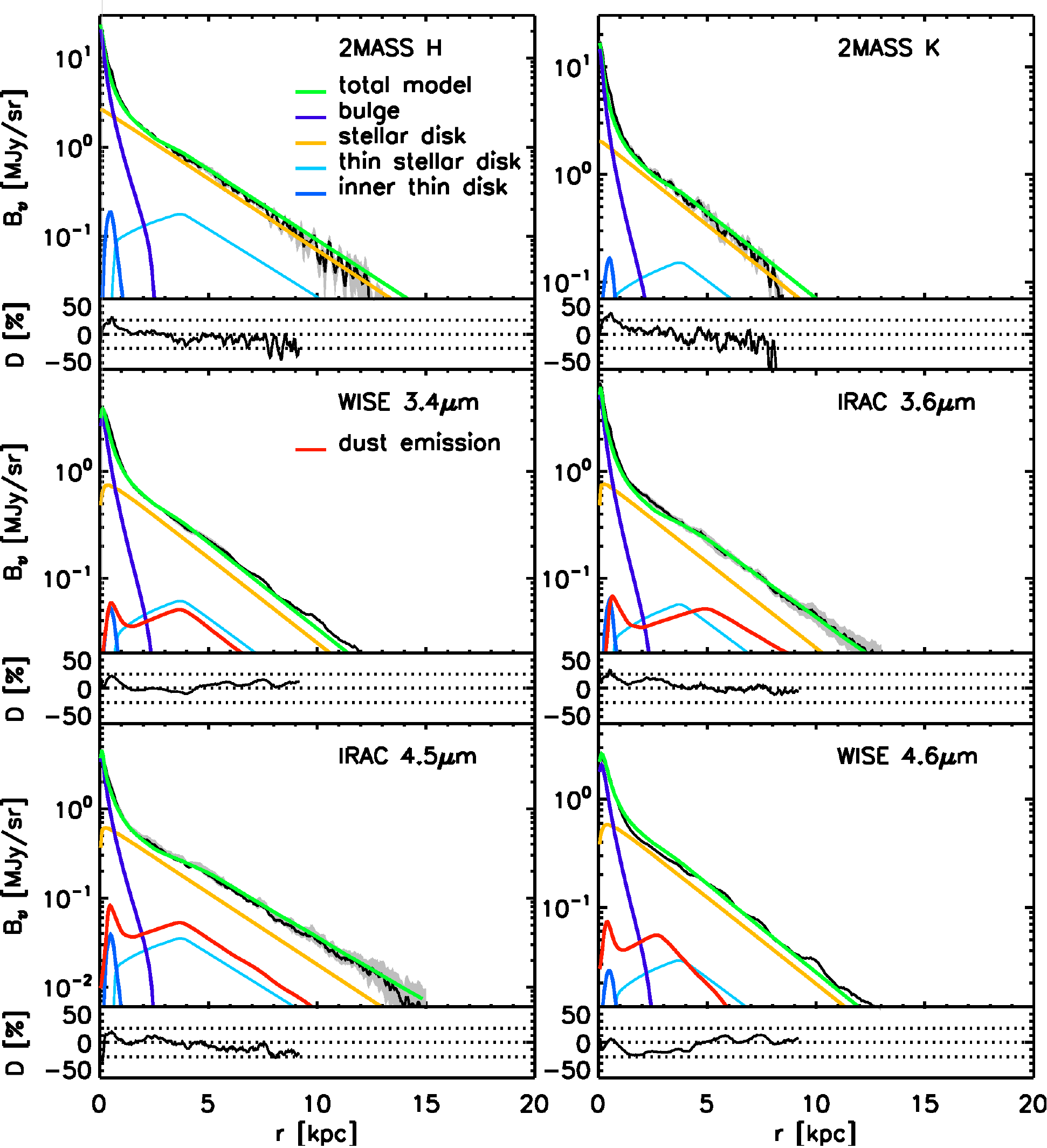}
\caption{As in Fig.~\ref{fig:UVopt_sbp}, but for the NIR profiles. At longer wavelengths a component of dust emission is also present, plotted with red lines.}
 \label{fig:NIR_sbp}
\end{figure*}

 \begin{figure*}
  \centering
  \includegraphics[width=0.8\linewidth]{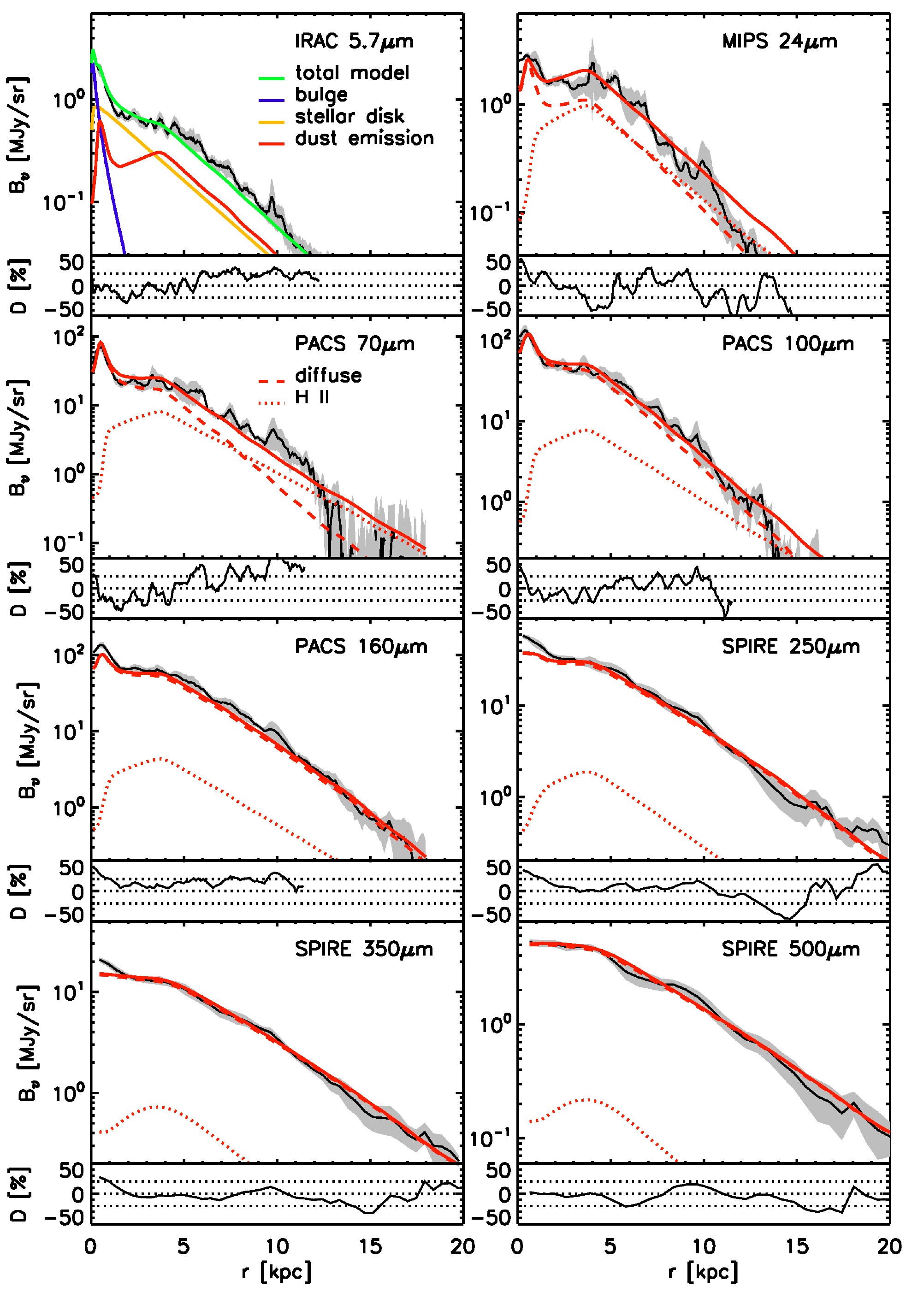}
\caption{As in Fig.~\ref{fig:UVopt_sbp}, but for the MIR/FIR/submm range. Except for the 5.7\,${\mu}$m, the emission is solely coming from dust. The plots also show the main components of dust emission: the diffuse (dashed-line) and the clumpy (dotted-line) component.}
 \label{fig:FIR_sbp}
\end{figure*}

\noindent
The errors in the azimuthally averaged profiles are also calculated according to Thirlwall et al. (2020)(their Appendix A). They include, in addition to the calibration errors, background fluctuations and Poisson noise,  a component of error due to the departure of the real structure from axi-symmetry, which we call  configuration noise. 

Examples of azimuthally averaged observed images can be seen in Fig.~\ref{fig:N628_Images} (middle column). The resulting observed profiles together with their errors are plotted in Figs.~\ref{fig:UVopt_sbp} -
\ref{fig:FIR_sbp}. 

\begin{figure*}
\centering
\begin{minipage}{.4\textwidth}
  \centering
  \includegraphics[width=1.0\linewidth]{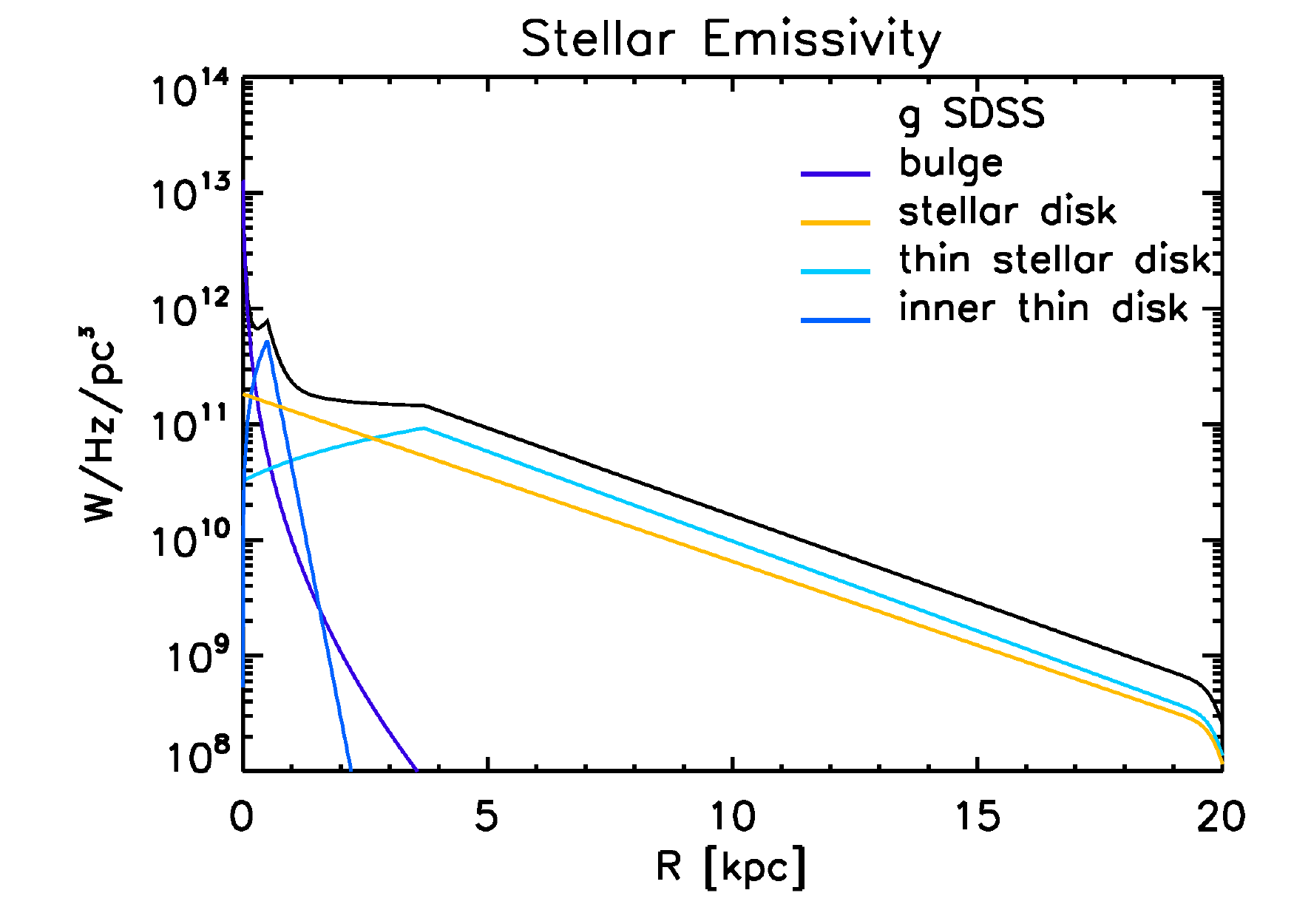}
\end{minipage}%
\begin{minipage}{.4\textwidth}
  \centering
  \includegraphics[width=1.0\linewidth]{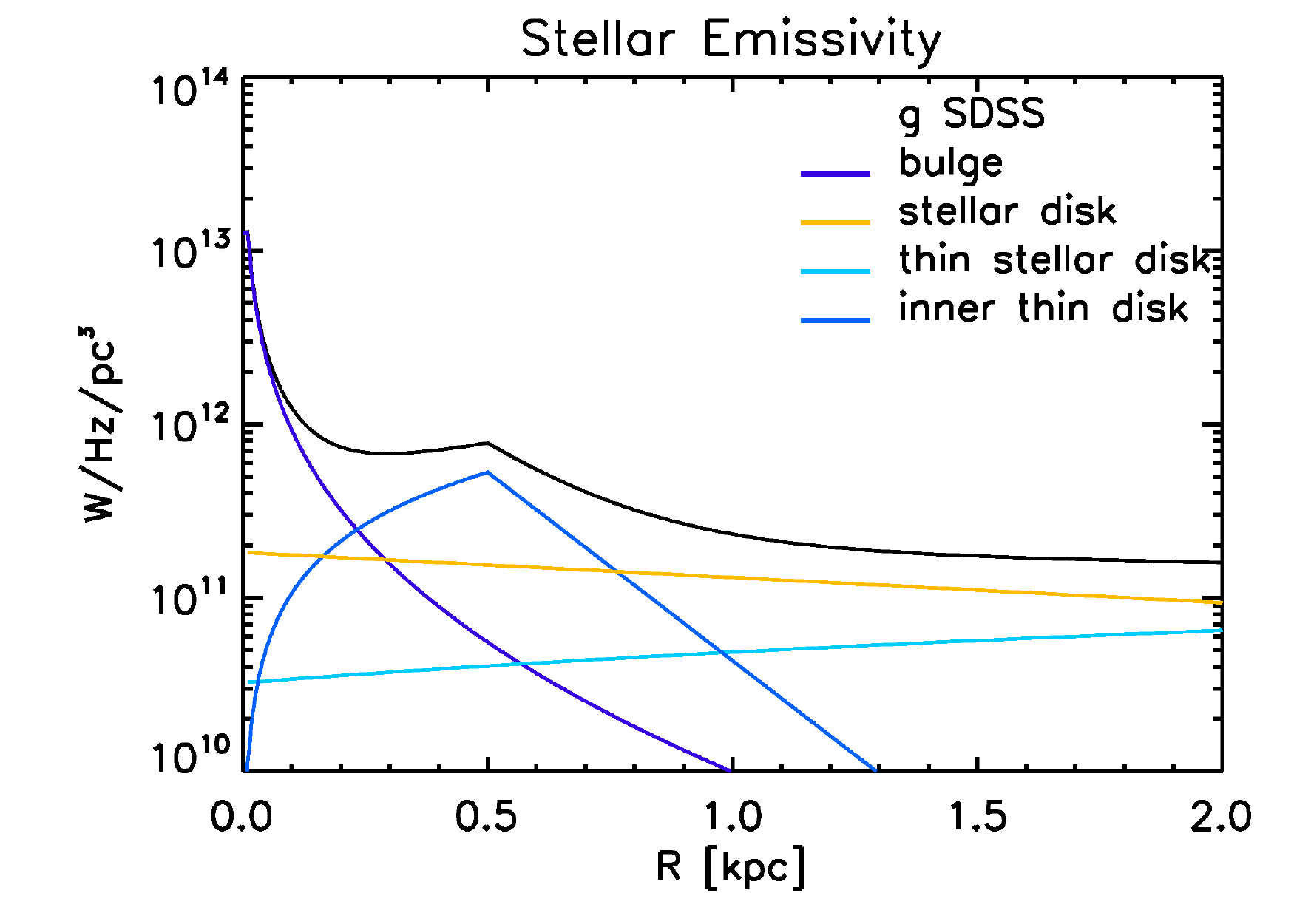}
\end{minipage}

\begin{minipage}{.4\textwidth}
  \centering
  \includegraphics[width=1.0\linewidth]{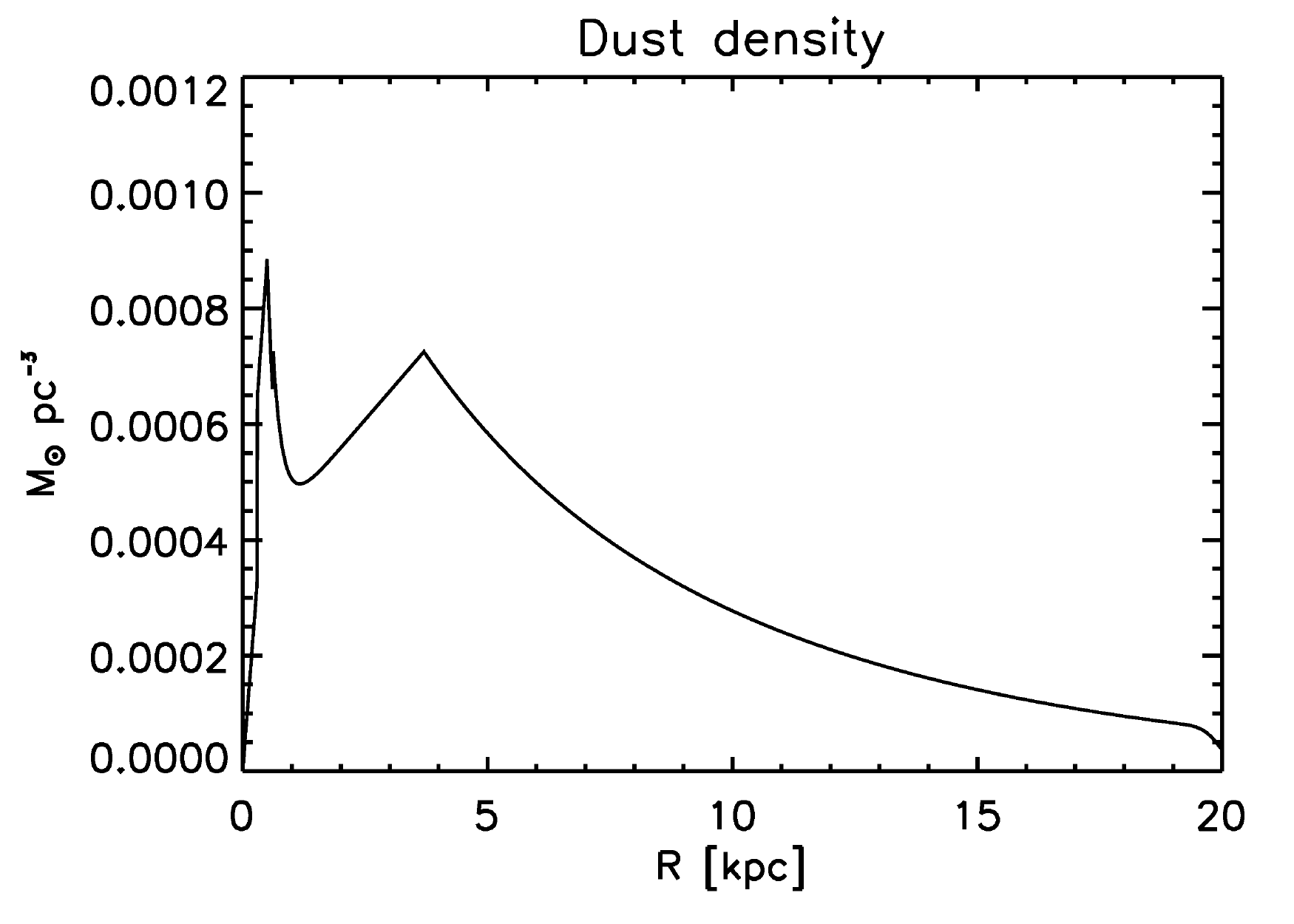}
\end{minipage}%
\caption{Top left: The radial profile  in the midplane ($z=0$) of the g-band stellar emissivity best fitting model. The coloured lines represent the model components for the bulge (purple), stellar disk (yellow), thin stellar disk (light blue) and inner thin stellar disk (dark blue). The total model is shown by the black continuous line. Top right: The same as the top left, but zoomed in the inner galaxy ($0-2$\,kpc). Bottom: The radial profile in the midplane ($z=0$) of the dust distribution for the best fitting model.}
\label{fig:emiss_opacity}
\end{figure*}

\section{Structural components used in the analysis}
\label{sec:struct}

The goal of this paper is to model NGC~628 with an axi-symmetric model. But how good is the approximation of axi-symmetry for this particular galaxy? To answer this question we looked at the appearance of the observed images (see left column of Fig.~\ref{fig:N628_Images}) and of the azimuthally averaged observed images (see middle column of Fig.~\ref{fig:N628_Images}). 
One can see that, despite the detail structure of the disk, the observed images resemble quite well the average images. One can also see that, when averaging, the images become rather smooth, with exponential like profiles. Exception is the UV image, which displays more ring like structures, although a larger binning of the annuli could also smooth the image. In other words the UV image has fluctuations centred on a declining exponential slope. We conclude that NGC~628 is well suited for azimuthal symmetry. As we will see later, this conclusion is consistent with the very good energy balance predicted between energy absorbed and re-emitted by dust, which accounts for the the good match between predicted quantities and data.

The galaxy structure used in our modelling approach is based on that used in the Popescu et al. (2011) (hereafter, PT11) axi-symmetric model. It comprises a dust-less bulge, a disk of young stars and associated dust (the thin disk), and a disk of old stars and associated dust (the disk). In this model, the thin stellar disk and associated dust is composed of diffuse and localised components, the latter being associated with the birth clouds of stars. To this standard model, we add a further disk of young stars and associated dust, accounting for a spike in the UV surface brightness profiles close to the centre of NGC~628 (see Fig.~\ref{fig:UVopt_sbp}). We refer to it as the \lq\lq inner thin disk\rq\rq.

The profiles of the stellar emissivity and dust density are given by analytical functions, taken from the generic model of PT11, with adaptations for NGC~628.
 In the inner disk, the radial distributions are linear, with a cutoff near the galaxy centre.
In the outer disk, the profiles follow the generic model, with exponentials in the radial direction.

The functions for the stellar volume emissivity and dust density in the disk configuration, $w_\nu(R, z)$, are of the form:

\begin{equation}
\label{eq:model}
w_\nu (R,z) = \begin{cases}

{\displaystyle
0
} \\
\hspace{4.7cm} {\rm if}\hspace{0.1cm} R < R_{\rm tin}\\\\
{\displaystyle
A_o\left[\frac{R}{R_{\rm in}}(1-\chi) + \chi\right]
\exp\left(-\frac{R_{\rm in}}{h_{\rm i}}\right) T}\\
\hspace{4.7cm} {\rm if}\hspace{0.1cm} R_{\rm tin} < R < R_{\rm in}\\\\
{\displaystyle
A_o
\exp{\left(-\frac{R}{h_{\rm i}}\right)T}
\hspace{1.6cm}{\rm if}\hspace{0.1cm}
R_{\rm in} \leq R \leq R_{\rm t}
}
\end{cases}
\end{equation}

\noindent with

\begin{equation}
\label{eq:chi}
\chi = \frac{w_\nu(0,z)}{w_\nu(R_{\rm in},z)}
\end{equation}

\noindent
assuming exponentials for the vertical distributions:
\begin{equation}
\label{eq:TRz}
T = \frac{z_{\rm i}(0)}{z_{\rm i}}\rm{exp}{\left(-\frac{z}{z_{\rm i}}\right)}
\end{equation}

\noindent
where
$R$ and $z$ are the radial and vertical coordinates, $h_{\rm i}$ is the scale-length, $z_{\rm i}$ is the scale-height, $A_{\rm o}$ is a scaling factor (the central emmissivity), $\chi$ characterises the slope of the linear distribution within the inner radius $R_{\rm in}$,
 $R_{\rm tin}$ is the inner truncation radius for the linear distributions, and $R_{\rm t}$ is the outer truncation
radius.

The bulge stellar emissivity $w_\nu(R, z)$ is represented by a Sersic function:

\begin{equation}
\label{eq:bulge}
w_\nu(R,z) = w_\nu(0,0)\,
\sqrt{\frac{b_{\rm s}}{2\pi}}\,\frac{(a/b)}{R_{\rm eff}}\,
\eta^{(1/2n_{\rm s})-1}
\exp{(-b_{\rm s}\, \eta^{1/n_{\rm  s}})}
\end{equation}
\noindent
with
\begin{equation}
\eta(R,z) = \frac{\sqrt{R^2 + z^2(a/b)^2}}{R_{\rm eff}}
\end{equation}
where $b/a$ is the axis ratio, $n_{\rm s}$ is the Sersic index, $R_{\rm eff}$ is the
effective radius and
$b_{\rm s}$ is a constant.

\subsection{The stellar disk}

The stellar disk is taken to contain old stellar populations, although it is not defined in terms of age of stars, but in terms of geometry. Thus, the disk contains all the stars that had time to migrate into a thicker disk (a few hundred parsec scale-height) than the original molecular layer from which the stars formed. 

The light from the stellar disk dominates the output in the optical and near-IR regions up to about $6\,\mu$m, since at even longer wavelengths it is the dust continuum that becomes  significant. At $3-5\,\mu$m, the dust attenuation is relatively low and we essentially observe the intrinsic direct light from the old stellar population, making it easier to constrain the model for the stellar emissivity at these wavelengths. Nonetheless, since the observational data in this region of the electromagnetic spectrum is obtained with IRAC/WISE, and since their bandpasses  are relatively wide, PAH emission may make a significant contribution to the fluxes at 3\,$\mu$m and 6\,$\mu$m. This is taken into account in the modelling approach. The intrinsic spectral luminosity densities of the stellar disk ($L_{\nu, {\rm s}}^{{\rm disk}}$) are free parameters, and are to be determined by matching the model to the surface photometry at each sampled optical/NIR wavelength. For this component we use the shortened form `disk' in the parameter superscripts.

The scale-length of the stellar  ($h_{\rm s}^{\rm disk}$) is a free parameter, and is also to be determined at each sampled wavelength in the $0.3-5\,\mu$m range. In this case a pure exponential was identified, meaning  $R^{\rm disk}_{\rm in, s}$=0 ($R^{\rm disk}_{\rm tin, s}$=0). The scale-height ($z_{\rm s}^{\rm disk}$) was fixed by assuming the same ratio to the B-band scale-length as in PT11, where the scale-length refers to a single exponential fit to the radial profile. Tests showed that the choice of scale-height had minimal impact on the fitting, as long we do not depart from our generic model in which older stars spread above the dust layer. This is true for all model constituents, again as long as the general characteristics of our generic model are not broken.

Thus, the free  parameters of the stellar disk  $L_{\nu, {\rm s}}^{{\rm disk}}$ and $h_{\rm s}^{\rm disk}$  are constrained from fitting the model to the optical/NIR surface brightness photometry. 

\subsection{Thin stellar disk}

The main characteristic of the thin stellar disk is, as the name indicates, its low scale-height. We assume that this disk contains the heavily attenuated young stellar population, although again it is defined in terms of geometry, and not in terms of stellar age. This disk is mainly observed in the UV and also contributes significantly to the bluest optical fluxes. This means that we have a direct constraint on the young stellar population, unlike the situation for an edge-on galaxy, where the young stars in the plane of the disk are completely obscured from our viewing perspective.

As in the case of the stellar disk, the spectral luminosity densities $L_{\nu, {\rm s}}^{\rm tdisk}$ are free parameters, and are found by matching the model to the surface brightness photometry at each sampled wavelength. For this component we use the shortened form `tdisk' in the parameter superscripts. 

The scale-height of this disk is assumed to be $z_{\rm s}^{\rm tdisk}=90$\,pc, i.e., the same as the value in PT11. 
 Following PT11, the values of the scale-height $z_{\rm s}^{\rm tdisk}$ and scale-length ($h_{\rm s}^{\rm tdisk}$) are assumed to be the same at all wavelengths. The inner radius $R^{\rm tdisk}_{\rm in, s}$ is a fixed parameter, easily identified from the shape of the observed profiles. The geometrical parameters $h_{\rm s}^{\rm tdisk}$, $\chi_{\rm s}^{\rm tdisk}$,  and $R^{\rm tdisk}_{\rm tin, s}$  are free parameters and are mainly constrained from fitting the model to the UV surface brightness photometry.

\subsection{Inner thin stellar disk}
\label{sec:stellar_ring}

The GALEX UV surface photometry shows a prominent spike in the inner galaxy, at around 0.5\,kpc from the centre (see Fig.~\ref{fig:UVopt_sbp}). This is also visible on the GALEX images as a bright region, curving around the centre. This emission complex is modeled with the inner thin stellar disk component. In the optical/near IR, the inner thin disk is inconspicuous, where it is presumably overwhelmed by the light from the bulge. Dust associated with this formation is clearly seen on IR images, especially at 24\,$\mu$m (Fig.~\ref{fig:FIR_sbp}) and even at PACS wavelengths (Fig.~\ref{fig:FIR_sbp}). This feature was detected in CO by Wakker et al. (1995) and James et al. (1995). It is a prominent component and a notable characteristic of the galaxy.

The spectral luminosity densities of the inner thin stellar disk, $L_{\nu, {\rm s}}^{\rm i-tdisk}$, in the UV are free parameters, and are constrained by matching the model to the surface photometry at each sampled wavelength. For this component we use the shortened form `i-tdisk' in the parameter superscripts. In the optical-near IR, the wavelength dependence of the inner thin disk's SED is assumed to be equal to that of the template from PT11 (see their Table~E.2).

The inner thin stellar disk model component is assumed to have the same scale-height as the thin stellar disk: $z_{\rm s}^{\rm i-tdisk}=z_{\rm s}^{\rm tdisk}$. The spike in the UV surface brightness profiles declines steeply toward the galaxy centre, but less steeply toward the outer disk (Fig.~\ref{fig:UVopt_sbp}). To create this profile in the model, we use a small scale-length for the inner thin disk ($h_{\rm s}^{\rm i-tdisk}$), set the inner radius $R_{\rm in, s}^{\rm i-tdisk}$ equal to the position of the peak, and $\chi_{\rm s}^{\rm i-tdisk}=0$ and $R_{\rm tin, s}^{\rm i-tdisk}=0$. 

Thus, the free parameters of the inner thin stellar disk are: 
$L_{\nu, {\rm s}}^{\rm i-tdisk}$, $h_{\rm s}^{\rm i-tdisk}$, $\chi_{\rm s}^{\rm i-tdisk}$ and $R_{\rm tin, s}^{\rm i-tdisk}$.  

\subsection{Stellar Bulge}

The stellar bulge dominates the $0.3-5\,\mu$m output from the inner galaxy. $R_{\rm eff}$ is a free parameter and  is determined at each sampled wavelength. Similarly, the intrinsic luminosity density of the bulge, $L_{\nu, {\rm s}}^{\rm bulge}$, is a free parameter and is also determined at each sampled wavelength.

\subsection{Dust disk}

The dust disk is in form of a diffuse component associated with the stellar disk. Its amplitude (face-on B-band opacity $\tau_B^f$) and geometrical parameters $h_{\rm d}^{\rm disk}$, $\chi_{\rm d}^{\rm disk}$, $R_{\rm in, d}^{\rm disk}$,  $R_{\rm tin, d}^{\rm disk}$ are mainly constrained from the submillimetre wavelengths.

In spiral galaxies, the scale-height of the dust disk was found to be about half that of the stellar disk, a general conclusion based on radiative transfer modelling of edge-on systems (see Xilouris et al. 1999; de Geyter et al. 2014), also adopted in PT11. We therefore use this ratio to the scale-height of the stellar disk to fix $z_{\rm d}$.

\subsection{Thin dust disk}

The thin dust disk is taken to represent the diffuse component associated with the spiral arms and the young stellar population. It has a smaller scale-length and a lower scale-height than the dust disk. As in PT11, the values are fixed to those of the thin stellar disk, i.e., $h_{\rm d}^{\rm tdisk}=h_{\rm s}^{\rm tdisk}$ and $z_{\rm d}^{\rm tdisk}=z_{\rm s}^{\rm tdisk}$.

\subsection{Inner thin dust disk}

This component was added to account for a relatively bright region surrounding the centre of the galaxy in the mid-IR images (Fig.~\ref{fig:FIR_sbp}). It is clearly associated with the bright feature seen at the same location in the UV images. We refer to this component as the \lq\lq inner thin dust disk\rq\rq\ . We assume $h_{\rm d}^{\rm i-tdisk}=h_{\rm s}^{\rm i-tdisk}$ and $z_{\rm d}^{\rm i-tdisk}=z_{\rm s}^{\rm i-tdisk}$.

\subsection{Clumpy component}

Optically thick clouds, in which massive stars are born, are an important component of the spatially integrated dust emission of star-forming galaxies.
The light from the young stars is locally absorbed  and remitted at intermediate wavelengths, where it can exceed the diffuse IR emission, despite the small filling factor of the SF regions in the galaxy. This is because the birth clouds are bathed in UV-dominated radiation from their stellar progeny in close proximity.
We assume that the clumpy component is distributed in the same geometry as the thin stellar disk. Its amplitude is defined by the F factor, the fraction of UV light locally absorbed by the birth clouds. This component is also known as the localised component, because the heating sources are close to the clouds.

\begin{table*}
\caption{The geometrical parameters of the model. All the scale parameters are in kpc.}
  \begin{tabular}{ | l | c }
    \hline
 \multicolumn{2}{c}{Free Parameters} \\ [-2ex]
 \multicolumn{2}{c}{}    \\
 \\
 $h_{\rm s}^{\rm disk}(ugrizJHKI_1I_2I_3$)  &                     (3, 2.9, 2.9, 2.8, 2.8, 2.7, 2.7, 2.7, 2.7,2.7, 2.7)$\pm10$\%                       \\
 $h_{\rm s}^{\rm tdisk}$                                 &   $2.8\pm0.2$     \\
 $h_{\rm d}^{\rm disk}$                                  &   $7.3\pm^{0.7}_{0.3}$\,    \\
 
 $h_{\rm s}^{\rm i-tdisk}$                                  &   $0.2\pm 0.05$            \\
 $\chi_{\rm d}^{\rm disk}$                              &   $0.8\pm0.1$           \\
 $\chi_{\rm s}^{\rm tdisk}$                              &   $0.35\pm0.1$   \\
 $\chi_{\rm d}^{\rm tdisk}$                              &   $0.6\pm0.1$      \\
 $R_{\rm tin, s}^{\rm disk}$                              &  $0.3\pm 0.05$           \\
 $R_{\rm tin, d}^{\rm disk}$                              &   $0.3\pm 0.05$    \\
 
 $R_{\rm tin, s}^{\rm tdisk}$                             &   $0.6\pm 0.05$     \\
 
 $R_{\rm eff}$                                           &   $0.82\pm0.05$              \\
 \multicolumn{2}{c}{}    \\ [-1ex]
 \multicolumn{2}{c}{Fixed Parameters} \\[-2ex]
 \multicolumn{2}{c}{}    \\
 \hline
 from data\\
 \hline
 $R_{\rm in, d}^{\rm disk}$                               &   $3.7$    \\
  $R_{\rm in, s}^{\rm disk}$                               &  $0.0$          \\
 $R_{\rm in, s}^{\rm tdisk}$                              &   $3.7$    \\
  $R_{\rm in, s}^{\rm i-tdisk}$                               &   0.5             \\
   $R_{\rm tin, s}^{\rm i-tdisk}$                              &   0.0              \\
   $R_{\rm t,s}^{\rm disk}$ & 20. \\
   $R_{\rm t,s}^{\rm tdisk}$ & 20. \\
   $R_{\rm t,d}^{\rm disk}$ & 20. \\
   $R_{\rm t,d}^{\rm tdisk}$ & 20. \\
     $\chi_{\rm s}^{\rm i-tdisk}$                              &   $0.0 $             \\
 $\chi_{\rm d}^{\rm i-tdisk}$                              &   $0.0$              \\
  $n_{\rm s}$                                             &   2                     \\
  \hline
  from model\\
  \hline
  $h_{\rm d}^{\rm tdisk}$                                 &  2.8     \\
 $R_{\rm in, d}^{\rm tdisk}$                              &   3.7      \\
 $R_{\rm tin, d}^{\rm tdisk}$                             &   0.6      \\
 $h_{\rm d}^{\rm i-tdisk}$                                  &   0.2              \\
 $R_{\rm in, d}^{\rm i-tdisk}$                               &   0.5              \\
 $R_{\rm tin, d}^{\rm i-tdisk}$                              &   0.0             \\
 $z_{\rm s}^{\rm disk}$        &  0.215                  \\
 $z_{\rm d}^{\rm disk}$        &  0.140                 \\
 $z_{\rm s}^{\rm tdisk}$        & 0.09                 \\
 $z_{\rm d}^{\rm tdisk}$        &  0.09                 \\
 $z_{\rm s}^{\rm i-tdisk}$        &   0.09                 \\
 $z_{\rm d}^{\rm i-tdisk}$        &   0.09                  \\ \hline
  \end{tabular}
  \label{tab:param} 
  \end{table*}

\section{Radiative Transfer Calculations}
\label{sec:rt}

We use the radiative transfer code of PT11, which is an adaptation of the code from  Kylafis \& Bahcall (1987). The code is based on a ray-tracing algorithm that calculates the intensity arriving at each pixel as a sum of direct and scattered light. This sum is dominated by the terms for the unscattered and first order scattered light, calculated accurately in the Kylafis \& Bahcall method. The contribution to the summed intensity from higher orders diminishes rapidly and is approximated without significant loss of accuracy. For these terms, the code uses the method of scattered intensities (see review by Kylafis \& Xilouris 2005).

We used higher sampling in the inner galaxy, where small scale structure is present, than in the outer disk. The sampling varied in the radial direction from 50\,pc in the centre to 2\,kpc in the outer disk. The sampling in the vertical direction varied from 50\,pc in the plane to 500\,pc in the outer halo. The models were produced with a truncation radius of $R_t=20$\,kpc.

We also used the Dartray code (Natale et al. 2014, 2015, 2017) for producing dust maps and radiation fields in the dust emission and to cross check the calculations.

The dust model consists of a mixture of silicates, graphites, and PAH molecules, which are thought to be responsible for sharp peaks in the mid-IR of star-forming galaxies. It is from Weingartner et al. (2001) and Draine \& Li (2007).

\section{Optimisation Procedure}
\label{sec:optim}

The free parameters of the model were derived by comparing the azimuthally averaged profiles obtained from the available imaging observations (see Sect.~\ref{sec:data}) with the corresponding model profiles.  Model profiles were derived from model images using the same surface-brightness photometry procedure as applied to observations. The whole fitting and optimisation procedure involved an iterative process whereby the RT calculations were run by changing only one or a pair of free parameters at a time. The optimisation strategy is to use specific wavelengths for constraining different geometrical and amplitude parameters, and making use of the fact that different parameters affect significantly only the emission at specific wavelengths (see PT11 for details).  This approach avoids unnecessary combinations of parameter values and allows for a solution to converge after a minimum number of iterations. 
The method is
similar to that used by Thirlwall et al. (2020), when fitting the imaging observations of M33.

To start we ran an RT calculation with an informed guess of the free parameters, obtained by fitting the spatially integrated spectral energy distribution (SED) with generic models from the library of PT11. We developed an algorithm, which searches the library, finds the best-fitting model, and explores the surrounding parameter space to arrive at a more precise solution. The generic model assumes only three stellar components: a stellar disk, a thin stellar disk, and a bulge (while our final solution for NGC~628 includes a 4th component - the inner thin stellar disk).
Intrinsic stellar SEDs in the generic model were approximated by spectral templates for typical young and old stellar populations (see PT11 for details).  

Initial estimates of the geometrical parameters were obtained by fitting the optical and near-IR images with the two-dimensional algorithm GALFIT (Peng et al. 2002). We found that the surface photometry could be matched with a Sersic component representing the bulge and an exponential component representing the stellar disk. The GALFIT derived Sersic index and effective radius of the bulge, and the scale-length of the disk, at the sampled optical wavelength, were inputs to the RT calculation.

Using the initial guess parameters we ran the RT code to produce initial model maps at all sampled wavelengths. These were then compared with observed maps. It was immediately apparent that the pure exponentials used in the generic model needed to be altered, by allowing for an inner radius and an inner linear slope (see Eqs.~\ref{eq:model} and \ref{eq:chi}). The inner radii were unambiguously derived from the observed profiles for all stellar and dust components, and were fixed throughout the subsequent optimisation procedure. This proceeded as follows:
\begin{enumerate}

\item We started by comparing the model and observed maps in the submm. In particular the SPIRE 500\,$\mu$m band is in a region dominated by thermal emission from the diffuse dust component. It is situated on the Rayleigh-Jeans tail of the emission, where the amplitude is essentially proportional to the dust opacity. The SPIRE 500\,$\mu$m band covers the longest sampled wavelength. It is the least sensitive to the luminosity of the heating sources and chosen to constrain the distribution and amplitude of the diffuse dust component.

Thus, by comparing model and observed 500\,$\mu$m profiles for different combinations of the parameters related to the diffuse dust, we were able to determine preliminary values for the amplitude parameters  $\tau_{\rm B}^{\rm f,tdisk}$ and $\tau_{\rm B}^{\rm f,disk}$ and the. 
geometrical parameters $h_{\rm d}^{\rm disk}$ and $\chi_{\rm d}^{\rm disk}$.
The inner truncation $R_{\rm tin, d}^{\rm disk}$ was unconstrained at this stage, because of the low resolution in the submillimetre. A need for an inner truncation was apparent at a later stage, when comparing the PACS profiles with the corresponding model ones.

\item GALEX traces the emission from the massive, short-lived stars and provided the constraint on the amplitude and geometry of the thin stellar disk and inner thin stellar disk. Since the inner thin stellar disk is most pronounced in the UV, it is only at this step in the optimisation procedure that it was added as a separate component. The UV stellar emissivities   were found by running RT calculations with the dust parameters from step 1, and by varying the relevant free parameters until a good match to the data is achieved. 

The amplitude parameters  were determined by scaling the model of escaping starlight to best match the surface photometry.
Thus, the intrinsic luminosity densities  $L_{\nu, {\rm s}}^{\rm tdisk}$ and $L_{\nu, {\rm s}}^{\rm i-tdisk}$ at the observed UV wavelengths are free parameters and are derived from the fitting procedure. At the other sampled UV wavelengths (where no observations exist) we scaled the model according to our initial SED template (for the young stellar population) and considering some interpolation to smooth the transition between wavelengths with and without observations.

Seven geometrical parameters were constrained in this second step. 
Of this three parameters  ($\chi_{\rm s}^{\rm i-tdisk}$, $R_{\rm tin}^{\rm i-tdisk}$ and $R_{\rm tin}^{\rm i-tdisk}$) were fixed from data, as explained in Sect.~\ref{sec:stellar_ring}. As such there are only four free parameters used in the optimisation: $h_{\rm s}^{\rm tdisk}$, $\chi_{\rm s}^{\rm tdisk}$, $R_{\rm tin, s}^{\rm tdisk}$, and $h_{\rm s}^{\rm i-tdisk}$.

\item Following constraints from PT11 we fixed the geometrical parameters of the thin dust disk and inner thin disk to be the same as those of the thin stellar disk and inner thin disk: $h_{\rm d}^{\rm tdisk}=h_{\rm s}^{\rm tdisk}$, $R_{\rm in, d}^{\rm tdisk}=R_{\rm in, s}^{\rm tdisk}$, $h_{\rm d}^{\rm i-tdisk}=h_{\rm s}^{\rm i-tdisk}$, $R_{\rm in, d}^{\rm i-tdisk}=R_{\rm in, s}^{\rm  i-tdisk}$,
$\chi_{\rm d}^{\rm i-tdisk}=\chi_{\rm s}^{\rm i-tdisk}$. 

\item With the parameters characterising the diffuse dust, thin stellar disk and inner thin disk fixed from the previous steps, we ran further RT calculations to derive the parameters of the stellar disk and bulge, and some of the parameters (amplitude) of the thin stellar disk. This was done by fitting the model to the corresponding surface photometry in the optical/NIR.

While the geometrical parameters of the thin stellar disk were derived in step (ii), the luminosity densities  in this wavelength range  were still to be found.
The thin stellar disk is  an important contributor to the bluest optical bands, but its contribution is uncertain at longer optical/NIR wavelengths, where the old stellar population dominates the output. Because of this, the thin stellar disk was only optimized in the $u\,g\,r$-bands, and at longer wavelengths its intrinsic SED  was assumed unchanged from the generic spectral template (see Table~E.2 in PT11). In the $u\,g\,r$-bands, the contribution of the thin stellar disk and the stellar disk can be disentangle because of their different geometry: the thin stellar disk has an inner radius, while the stellar disk seems to be represented by a pure exponential extending right to the centre (no inner radius). In particular the amplitude of the thin stellar disk was constrained from a dip in the observed profile at smaller radii (between 1-2\,kpc).

Similarly, while the geometrical parameters of the inner thin stellar disk are derived in step (ii) in the UV, the luminosity densities in the optical/NIR were still to be found. Because in this wavelengths range  the bulge dominates the output from the inner galaxy, the optical/NIR emission from the inner thin stellar disk is assumed to have the same wavelength dependence as the generic spectral template used for the young stellar population (see Table~E.2 in PT11), and the overall scaling is determined from the UV.

Here we should also mention that the initial guess value for the Sersic index $n_{\rm s}$ of the bulge was 1 and was derived from the GALFIT procedure. However, when comparing the observed and model profiles it became apparent that a slightly larger value for $n_{\rm s}$ would do a better job. 
Because of this we considered the value $n_{\rm s} =2$, which provided a very good match to the data. We adopted this value at all wavelengths without any need for iterations. As such we consider the parameter $n_{\rm s}$ fixed rather than a free parameter, in the sense that we did not have to iterate on this during the optimisation. 

Thus, the fitting of the model to the optical-NIR surface photometry resulted in constraining the following parameters: 
 the intrinsic luminosity density $L_{\nu, {\rm s}}^{\rm disk}$, $L_{\nu, {\rm s}}^{\rm bulge}$, and $L_{\nu, {\rm s}}^{\rm tdisk}(ugr)$, the scale-length $h_{\rm s}^{\rm disk}$ and bulge effective radius $R_{\rm eff}$.

\item The localised dust emission within the star-forming complexes is a significant contributor to the spatially integrated SED over a relatively narrow (20-60\,$\mu$m) region. At 24\,$\mu$m, it is the dominant component. The next step used the MIPS 24\,$\mu$m surface photometry to constrain the amplitude of the clumpy component. The amplitude of the diffuse dust component was fixed in previous steps, leaving us free to determine the amplitude of the localised emission, defined by the F factor. 
\item With most of the parameters fixed from the previous steps we compared the model and observed profiles at all sampled wavelengths, including the 70-160\,${\mu}$m regime, and made small adjustments to the parameters, such that the best fit for the whole UV/optical/FIR/submm SED is obtained.
\item Colour corrections are needed to convert monochromatic flux densities to the actual flux densities at the reference wavelengths (column~1 of Table~\ref{obs}). This necessity arises from the flux calibration of the data products assuming that the source continuum is either stellar-like, or flat. Colour corrections are essential for a star-forming galaxy. They can be significant on the Wien-side of the dust emission ($20-70\,\mu$m) and where emission from Polycyclic Aromatic Hydrocarbon (PAH) molecules contributes to the fluxes ($3-20$\,$\mu$m). The colour corrections were calculated by feeding our model SED into the filter response curves. 
Corrections were made to the observed fluxes, requiring small adjustments to the best-fitting model.

\end{enumerate}

The error analysis of the model parameters follows Thirlwall et al. (2020) and the resulting geometric parameters and their corresponding errors are listed in Table~\ref{tab:param}. The resulting amplitude parameters of the stellar components (the intrinsic spectral luminosity densities) are listed in Table~\ref{tab:intrlum}.

The quality of the fits was judged by eye for a preliminary solution, and then a ${\rm chi}^2$ analysis was used to explore the surrounding parameter space. The value of ${\rm chi}^2$ is given by:

\begin{equation}
{\rm chi}^2_{\rm j}=\sum_{\rm i} \frac{(B^{\rm obs}_{\rm ij} - B^{\rm mod}_{\rm ij})^2}{\sigma^2_{\rm ij}},
\end{equation}
where $B^{\rm obs}_{\rm ij,r_i}$ is the average observed brightness in annulus $i$ at waveband $j$, $B^{\rm mod}_{\rm ij}$ is the average model brightness in annulus $i$ at waveband $j$, and $\sigma_{\rm ij}$ is the standard deviation of the brightness in annulus $i$. The reduced $\chi^2_{\rm r}$ at each wavelength is:
\begin{equation}
{\rm chi}^2_{\rm r} = \frac{{\rm chi}^2_{\rm j}}{N_j-M_j},
\end{equation}
where $N_j$ is the total number of annuli from the centre to the edge of the galaxy at the waveband $j$ and $M_j$ is the number of free parameters at waveband $j$, with $N_j>>M_j$. In Table~\ref{tab:chi} we list the ${\rm chi}^2_{\rm r}$ values for a few key wavelengths used to optimise the model. As one can see,  the ${\rm chi}^2_{\rm r}$ are reasonable, except for the $24\,{\mu}$m. At this wavelength the fit is still very good (see Fig.~\ref{fig:FIR_sbp}) as regards the overall luminosity and slope of the radial profile, but the observed profile is, as expected,  less smooth, since at this wavelength the star-forming clouds have a strong contribution. So the higher value of ${\rm chi^2_r}$ at this wavelength reflects the fact that the assumption of axi-symmetry is less good at this wavelength, rather than a problem with the overall energy balance. The very small value of ${\rm chi^2_r}$ in the 2MASS K-band is not due to an over-fit of the data, but rather to the very large background fluctuations, which increase the overall assumed errors. In addition, in this range, the data is very smooth, very similar to the axi-symmetric model. So the model functions are naturally very good matches to the data.

The reduced ${\rm chi^2_r}$ for the model across all wavelengths is
\begin{equation}
{\rm chi}^2_{\rm r}=\frac{\sum_{\rm j=0}^{\rm j=L}{\rm chi}^2_{\rm j}}{\sum_{j=0}^{j=L} (N_j-M_j)}.
\end{equation}
where $L$ is the total number of wavelengths with available observations. We obtained a total 
${\rm chi}^2_{\rm r}=1.89$.

\begin{table}

\caption{The values of ${\rm chi^2_r}$ for the best-fitting model.}
  \begin{tabular}{ l c}
    \hline
Band & ${\rm chi^2_r}$  \\
\hline
 GALEX NUV & 1.14 \\
 2MASS K & 0.02 \\
 MIPS 24 & 8.66 \\
 SPIRE 500 & 0.45 \\
 \hline
 Total & 1.89 \\
 \hline 
 \end{tabular}
 \label{tab:chi}
 \end{table}

\section{Results}
\label{sec:results}
\subsection{Model fits}

Figs.~\ref{fig:UVopt_sbp}-\ref{fig:FIR_sbp} show the NGC~628 UV-submillimetre azimuthally averaged surface brightness profiles in order of the effective wavelength of the band, from $0.15\,\mu$m to $500\,\mu$m. In each panel, the model, and its components, are superimposed on the corresponding observations as coloured lines. The residuals are shown below the plots, with dotted lines marking the $\pm20$\% level. Good fits to the multiwavelength profiles are obtained, with average residuals of 12\%. The corresponding solution for the intrinsic stellar emissivity in the g band and for the dust distribution is shown in Fig.~\ref{fig:emiss_opacity}.

\subsubsection{UV profiles}

The fits to the UV surface brightness  profiles are shown in Fig.~\ref{fig:UVopt_sbp}. The structure in the data profiles is due to the spiral arms and H\,{\sc ii} regions (see top left of Fig.~\ref{fig:N628_Images}). The model curves are  smooth because of the assumption of axisymmetry.  Between 13 and 20 kpc the radial profiles seem to indicate an outer faint disk that, at the resolution of the data, is still well blended into the main disk. Because of this we did not utilise an explicit different morphological component to account for this emission.

The spike, seen in the profiles at a radial distance of 0.5\,kpc, dominates the inner galaxy. It is a significant source of UV radiation and heating, 
and is well matched by the inner thin stellar disk model component (and associated inner thin dust disk).
The other component of the model shown in the figure is the young stellar disk (light blue curve). 

\subsubsection{Optical profiles}

The stellar bulge dominates the $0.3-5\,\mu$m profiles of the inner galaxy (Figs.~\ref{fig:UVopt_sbp} and \ref{fig:NIR_sbp}). At $0.35\,\mu$m and $0.47\,\mu$m, there is extra emission that is well matched with contribution from the inner thin stellar disk model component. In the bluest optical bands the thin stellar disk and the stellar disk provide comparable amounts to the total emission at around $R_{\rm in}^{\rm disk}$. In the $ugr$ bands the amplitude of the young stellar disk was constrained from a dip in the observed profile at smaller radii, between $1-2$\,kpc. It is reproduced by a combination of a pure exponential for the stellar disk ($R_{\rm in, s}^{\rm disk}=0$) and a thin stellar disk that has an inner radius, a shallow linear slope and a sharp inner truncation radius. The dip is shallow and absent at longer wavelengths. 

\subsubsection{NIR profiles}

Model fits to the IRAC and WISE surface brightness profiles are shown in Figs.~\ref{fig:NIR_sbp} and \ref{fig:FIR_sbp}. The contribution from the dust emission is significant beyond $3\,\mu$m, dominating at the longest IRAC/WISE wavebands. It is shown by the red line in the $3.40-5.80\,\mu$m profiles. 

In the galaxy image at $5.8\,\mu$m (not shown in this paper) 
the emission follows the spiral arms and HII regions, resembling those from GALEX. This suggests that the emission is dominated by PAHs, which trace the UV radiation field of young stars. Since the intensity of PAH emission strongly depends on PAH abundance, it is perhaps not surprising that our model, which has a fixed abundance, fails to account for the overall amplitude at $5.8\,\mu$m. Thus, the model under-predicts the observations, with D > $+20\% $ across the radial profile.

\subsubsection{MIR profiles}

The MIPS $24\,\mu$m profiles show several peaks from the hot dust surrounding the HII regions, including a large peak modelled with the inner thin dust disk model component. (Fig.~\ref{fig:FIR_sbp}). The emission at this wavelength  is a key measurement because the localised component makes a significant contribution in this band  (48\% at $R_{\rm in}^{\rm disk}$; see Fig.~\ref{fig:FIR_sbp}).  An important feature of the profiles, well predicted by our model,  is the fact that they have smaller extent than the profiles at longer FIR wavelengths. This is predicted to arise because of two factors. Firstly, the HII component, which accounts for half of the emission,  has in our model the same scale-length as the thin stellar disk, which was found to be much smaller than that of the dust disk shaping the emission at longer wavelengths. Secondly, the diffuse emission, which accounts for the other half, strongly traces dominant heating sources, which are the young stars from the thin stellar disk.  As such, both components of emission are predicted to have the same small scale-length, which is exactly what is seen in observations.

\subsubsection{FIR profiles}

The PACS $70-160\,\mu$m radial profiles from Fig.~\ref{fig:FIR_sbp} show a peak at 0.5\,kpc associated with the inner thin disk component. The peak in the profiles is followed by a plateau and a steeper decrease beyond $R_{\rm in}^{\rm disk}$. Our model is successful in reproducing these features, and in particular the prominence of the peak.  At these wavelengths, where the emission is strongly modulated by the heating sources, the peak at 0.5\,kpc is predicted to come from two dust components: the inner thin dust disk and the dust disk. This is because the dominant source of heating at the position of the peak - the inner thin stellar disk model component -   powers not only the dust from the inner thin disk itself, but also the underlying (and dominant) dust disk, enhancing the peak at $R_{\rm in}^{\rm i-tdisk}$, an effect most pronounced at $70\,\mu$m.

\subsubsection{Submm profiles}

The SPIRE $250-500\,\mu$m surface brightness profiles are shown in Fig.~\ref{fig:FIR_sbp}. At these wavelengths the peak modelled with the inner thin disk disappears, blending in a sort of
plateau extending throughout the central region and out to the $R_{\rm in}^{\rm disk}$. Our model is successful
 in reproducing this feature, which naturally occurs because of the small predicted opacity
 of the inner thin dust disk model component. In the submillimetre, the profiles are mainly shaped by dust opacity, and, because of this, they mainly trace the dominant opacity component, which is predicted to be the dust disk.
\begin{figure*}
\centering
\includegraphics[width=0.8\textwidth,angle=-0]{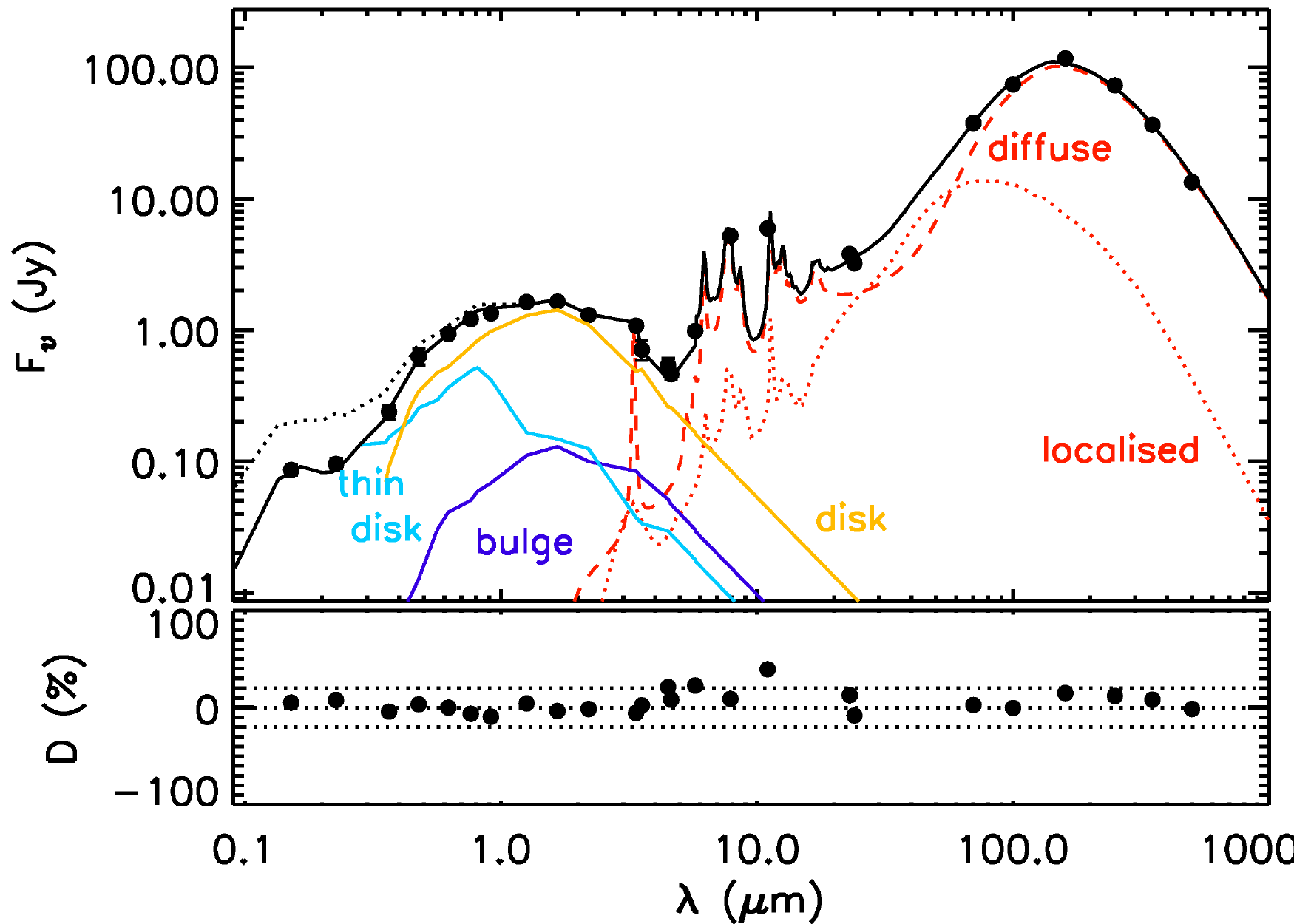}
\caption{Upper panel: SED of NGC~628, with the black squares representing the flux densities derived from observations, as listed in column (4) of Table~\ref{obs}. The best-fitting model is shown by the black continuous line. Its components are shown by the coloured lines: the purple line is the stellar bulge, the blue line is the young stellar disk, the orange line is the old stellar disk, the red dashed line is the diffuse dust emission and the red dotted line is the localised (clumpy) component. The black dotted lines shows the intrinsic (dust corrected) stellar SED. Lower panel: The residuals (D) shown as a percentage difference of the data and model. }
\label{fig:sed}
\end{figure*}

\subsubsection{The scale-length of the dust emission}

One interesting feature of the observed dust emission radial profiles is that the slope of the  emission outwards of the inner radius of the thin stellar disk, $R_{\rm in, s}^{\rm tdisk}=3.7\,$kpc,  becomes steeper with decreasing infrared wavelength. If we fit an exponential to this slope, we get a scale-length of 4.4\,kpc at 500\,${\mu}$m, but only 2.5\,kpc at 70\, ${\mu}$m.  This decrease by a factor of 1.76,  which is due to a combination of temperature gradients in the diffuse dust component and to the different scale-lengths of dust and heating sources, is accurately predicted by our model. Visually, this can be also noticed in Fig.~\ref{fig:N628_Images}, as the 500\,${\mu}$m image appears more extended than the 160\,${\mu}$m image, both in the data and in the model. Taken into account that the emission in the 70-350\,$\mu$m range is mainly predicted, and not fitted, we conclude that the spatial distribution of both the dust and the heating sources is correctly predicted by our model. We regard this as a great success of our model.

\subsection{Global SED}
\label{subsec:global_sed}

In the previous subsection we described the main features of our model fits to  the azimuthally averaged surface brightnesses. While this gives the most detailed information on the characteristics of the fit, it is of interest to 
also integrate the emission over the whole galaxy, and discuss the spatially integrated SED of NGC~628 with respect to our model. This is shown in Fig.~\ref{fig:sed}, with observed fluxes from column (4) of Table~\ref{obs}. The best fitting model is shown by the black line. Model stellar components are shown for the stellar bulge (purple line),  stellar disk (orange line) and thin stellar disk (light blue line). The model dust components shown are the localised (dotted red line) and diffuse emission (dashed red line). The panel below the SED shows the residuals and the dotted lines are at $\pm20$\%. The residuals are less than 20\% in most bands, averaging 8.6\%. The largest discrepancies are seen at IRAC 4 ($8\mu$m) and WISE W3 ($11\mu$m), and to a lesser extent at IRAC 3 ($5.7\mu$m), but these are complicated by emission from PAH molecules. The average residual is 6.7\% without these bands. For a radiative transfer model this is an excellent fit to the data. In particular the energy balance between absorbed and re-radiated light is very well matched.

The intrinsic stellar SED (as it would be observed in the absence of dust) is also shown in Fig.~\ref{fig:sed} with dotted black line. The difference between the apparent and intrinsic stellar SED is due to dust attenuation in the star-forming clouds and to dust attenuation in the diffuse component. The combined effect of the two components of attenuation present in our model results in a derived intrinsic SED that has a much shallower decrease towards shorter UV wavelengths than the apparent SED. In the optical/NIR bands dust attenuation becomes negligible.

\begin{figure}
\includegraphics[width=0.4\textwidth,angle=-0]{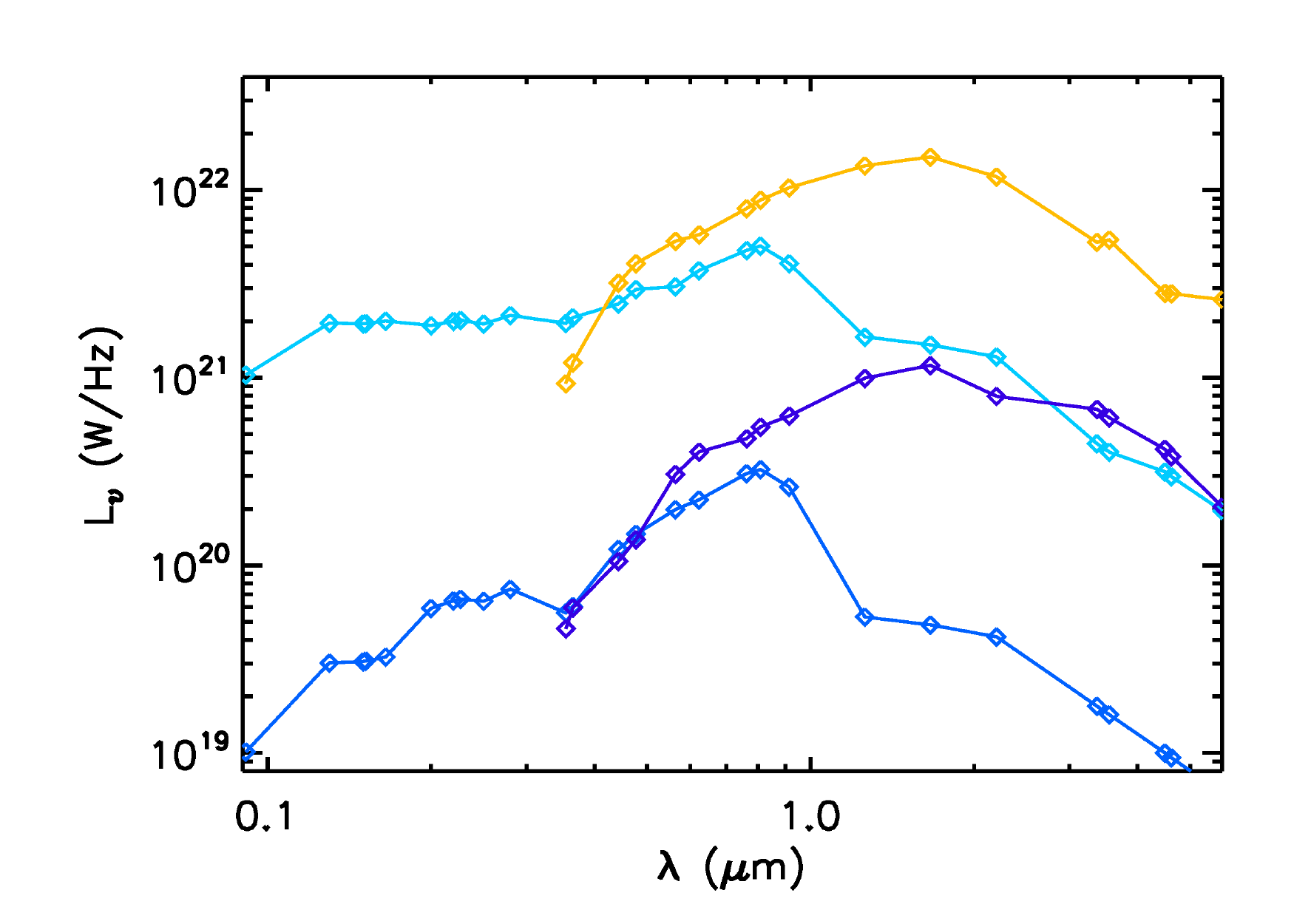}
\caption{The intrinsic SEDs of the stellar components in the model of NGC~628. The orange line represents the  stellar disk, the purple line the bulge, the light blue line the thin stellar disk and the dark blue line represents the inner thin stellar disk. The symbols show the sampled wavelengths of the model.}
\label{fig:intrinsic}
\end{figure}

The intrinsic SEDs of the individual stellar components are shown in Fig.~\ref{fig:intrinsic}. 
If we add up the output from all the stellar components and integrate over the spectral range we obtain an intrinsic stellar luminosity of $12.8\times 10^{36}$\,W ($3.34\times10^{10}$\,L$_{\odot}$). The luminosities of the different stellar components of the model are listed in  Table~\ref{tab:lumin_stellar}.  The inner thin stellar disk contributes only around $2\%$ to this luminosity. The old stellar population (from the stellar disk and bulge) makes up $46\%$ of the intrinsic stellar output, powering $29\%$ of the dust emission. This means that the dust emission is mainly powered by the young stellar populations (with 71$\%$). This is consistent with our previous
results obtained from RT modelling: $69\%$ for NGC~891 (Popescu et al. 2011), $80\%$ for M~33 (Thirlwall et al. 2020) and $71\%$ for the Milky Way (Natale et al. 2021). 
This fraction is larger than that derived in Nersesian et al. (2019) for galaxies of similar Hubble type, although the definition of young and old stars is quite different from our study, and  Nersesian's study does not involve RT methods. Studies that do involve RT methods find a broad range of values for this fraction on other galaxies: (63 per cent for M51 in de Looze et al. 2014, between 60 to 80 per cent for M33 in Williams et al. 2019, 50.2 per cent for M81 in Verstocken et al. 2020, 83 per cent for NGC 1068 in Viaene et al. 2020, 59 per cent for a sample of 4 barred galaxies in Nersesian et al. 2020a and 71.2 per cent for M51 in Nersesian et al. 2020b).

The dust luminosities derived by our model (total, diffuse and clumpy) are listed in Table~\ref{tab:lumin_dust}. The total dust luminosity is $5.03\times 10^{36}$\,W ($1.31\times10^{10}$\,L$_{\odot}$). This is comparable with the value of $4.72 \times 10^{36}\,$W derived by Aniano et al. (2012)  from a global fit to the dust emission SED. Here we note that Aniano et al. used a distance to NGC~628 of 7.2 Mpc, so the quoted value above has been scaled by us to account for the difference in the distance estimate. A larger value of $6.73 \times 10^{36}$\,W has been derived\footnote{The value has been adjusted by us to account for the difference in distance estimate (7.3 Mpc in Dale et al. 2009).} by Dale et al. (2009), but this was calculated using the simple formula for the total infrared luminosity ${\rm TIR}$ that linearly combines the 24, 70 and 160\,${\mu}$m, which cannot provide the same level of accuracy as a SED fit.

The dust emission SED is dominated by the diffuse component (see Fig.~\ref{fig:sed}), which makes up $79\%$$\pm 2\%$ of the total emission. The SED is typical for a spiral galaxy, peaking at around $160\,{\mu}$m and having a temperature of $\sim 18$\,K. We find that $39\%$ of the stellar light is absorbed by dust and re-radiated as dust emission. This matches with the overall estimates of this fraction for spiral galaxies in the local Universe (e.g. Popescu \& Tuffs 2002, Viaene et al. 2016), meaning that NGC~628 is an average galaxy with respect to this statistics.

\subsection{Star formation rate}

To calculate the SFR we use the derived intrinsic UV luminosity of NGC~628 and the calibration from Eq. 17 of Popescu et al. (2011). We find ${\rm SFR=(2.0\pm0.15)\,M_{\odot}yr^{-1}}$ and a SFR surface density of 
${\rm \Sigma_{\rm SFR}}=(1.6\pm0.2) 10^{-3}\,{\rm M}_{\odot}\,{\rm yr}^{-1}$\,kpc$^{-2}$ within 20\,kpc radius. ${\rm \Sigma_{\rm SFR}}$ is of course not constant over the surface of the disk, but, as we will see in Sect.~\ref{sec:sfr_distrib}, has a peak at 0.5\,kpc, which, in the model, corresponds to the position of the inner thin disk. Indeed, the inner thin disk is a site of enhanced star forming activity. Its SFR and SFR surface density are ${\rm SFR=0.04\,M_{\odot}yr^{-1}}$ and 
${\rm \Sigma_{\rm SFR}}=0.013\,{\rm M}_{\odot}\,{\rm yr}^{-1}$\,kpc$^{-2}$, 
respectively.

Our value for the total SFR is by far smaller than that derived in Calzetti et al. (2015) of ${\rm SFR}=3.38\pm 0.2\, {\rm M}_{\odot}\,{\rm yr}^{-1}$ (\footnote{Value adjusted for a distance of 9.5\,Mpc; $D=9.9$\,Mpc used in Calzetti et al (2015)}.). Calzetti et al. used the GALEX FUV data corrected for dust attenuation using the formalism from Lee et al. (2009). This  utilises the TIR/FUV ratio, where the TIR value is derived using the Dale et al. (2009) formula. We have already seen in the previous subsection that the TIR from Dale et al. overestimates the total dust emission by $34\%$. This translates in an overestimation of the A(FUV), as derived from eq. 6 of Lee et al. (2009)  by $28\%$. Of course the correlation between A(FUV) and the ${\rm TIR/FUV}$ has itself a large scatter, so perhaps it is not surprising to see a large discrepancy between these determinations.

A smaller value of ${\rm SFR}=2.4\,{\rm M}_{\odot}\,{\rm yr}^{-1}$ was derived in Sanchez et al. (2011) using integral field spectroscopy. This value is closer to our derivation, and possible consistent, although no error is attached to this determination.

More recently,  Enia et al. (2020) and Morselli et al. (2020) derived ${\rm SFR}=1.67\pm 0.41\,{\rm M}_{\odot}\,{\rm yr}^{-1}$ and ${\rm SFR}=1.56\pm 0.41\,{\rm M}_{\odot}\,{\rm yr}^{-1}$ (\footnote{Values adjusted for a distance of 9.5\,Mpc; $D=10.14$\,Mpc used in Enia et al. (2020) and Morselli et al. (2020).}) using the DustPedia  database (Davies et al. 2017) and a linear combination between the $L_{\rm FUV}$ and $L_{\rm IR}$ to calibrate the SFR (Eq.~1 from Morselli et al., and the calibration from Bell \& Kennicutt 2001 and Kennicutt 1998), with  $L_{\rm FUV}$ and $L_{\rm IR}$ derived from an SED fit. Similar values, of  ${\rm SFR}=1.6\pm 0.80\,{\rm M}_{\odot}\,{\rm yr}^{-1}$(\footnote{Values adjusted for a distance of 9.5\,Mpc; $D=9.8$\,Mpc used in Leroy et al. (2019)}), were derived by Leroy et al. (2019), using the Multiwavelength Galaxy Synthesis database and a combination of FUV+WISE4 data. All these recent values seem consistent with our derivation.

If we adopt the stellar mass $M_*=1.21\times10^{10}\,{\rm M}_\odot$ (\footnote{As before, the value is adjusted for a distance of 9.5\,Mpc.}) from Morselli et al. (2020) we obtain a specific star-formation rate ${\rm sSFR}=1.65\times 10^{-10}\,{\rm yr}^{-1}.$
If we adopt the stellar mass $M_*=1.64\times10^{10}\,{\rm M}_\odot$ (\footnote{As before, the value is adjusted for a distance of 9.5\,Mpc.}) from Leroy et al. (2019) we obtain a specific star-formation rate ${\rm sSFR}=1.22\times 10^{-10}\,{\rm yr}^{-1}.$

\begin{table}

\caption{The face-on dust opacity, $\tau_{\rm B}^{\rm f}$, of the inner thin dust disk, dust disk, and thin dust disk, at the inner radius of the inner thin dust disk and dust disk (thin dust disk) . The total dust opacity at the corresponding radii is also given in the last row.}
  \begin{tabular}{ l l  l l}
    \hline
 \multicolumn{4}{}{} \\ [-2ex]
$R_{\rm in}$ [kpc] & 0.5  & 3.7 &    \\ 
 \hline\\
 & $\tau^f_B$ & $\tau^f_B$\\
 inner thin disk & 0.82 & 0.00  & \\
 disk & 1.05 & 1.27 & \\ thin disk & 0.00 & 0.20&
 \\
 \hline\\
 total & 1.87 & 1.47 & \\
 \hline\\
 
 \end{tabular}
 \label{tab:tau}
 \end{table}

\subsection{Dust mass and opacity}

The total dust opacity is 
$\tau_{\rm B}^{\rm f}=1.47\pm0.12$  
at $R_{\rm in}^{\rm disk}$, with most of the opacity provided by the dust disk (see Table~\ref{tab:tau}). The maximum opacity occurs at the position of the inner thin disk, where there is a sharp peak, with $\tau_{\rm B}^{\rm f}=1.87$.
The galaxy is marginally optically thick at the centre, with $\tau^f_B(0)=1.16$, and optically thin beyond 6\,kpc. 

Using Eq.~B1-B6 from Thirlwall et al. (2020) we derive a dust mass of $M_{\rm dust}=(7.47\pm0.30)\times10^7\,{\rm M_\odot}$. If we consider a HI mass $M ({\rm HI}) = (6.44 \pm 0.2) \times 10^9 M_{\odot}$ (Walter et al. 2008) (adjusted) and a molecular gas mass  $M_{{\rm H2}}=(4.35\pm0.6)\times 10^9 M_{\odot}$ (Aniano et al.)(adjusted), then the total gas mass is $(10.8\pm 0.63) \times 10^{9}M_{\odot}$, giving a dust-to-gas ratio of $D/G=0.0069\pm 0.0005$. Within errors this is comparable to $D/G=0.0082\pm0.0017$ from Aniano et al. (2012).

\subsection{Radiation Fields}

\begin{figure}
 \centering
\includegraphics[width=1.0\linewidth]{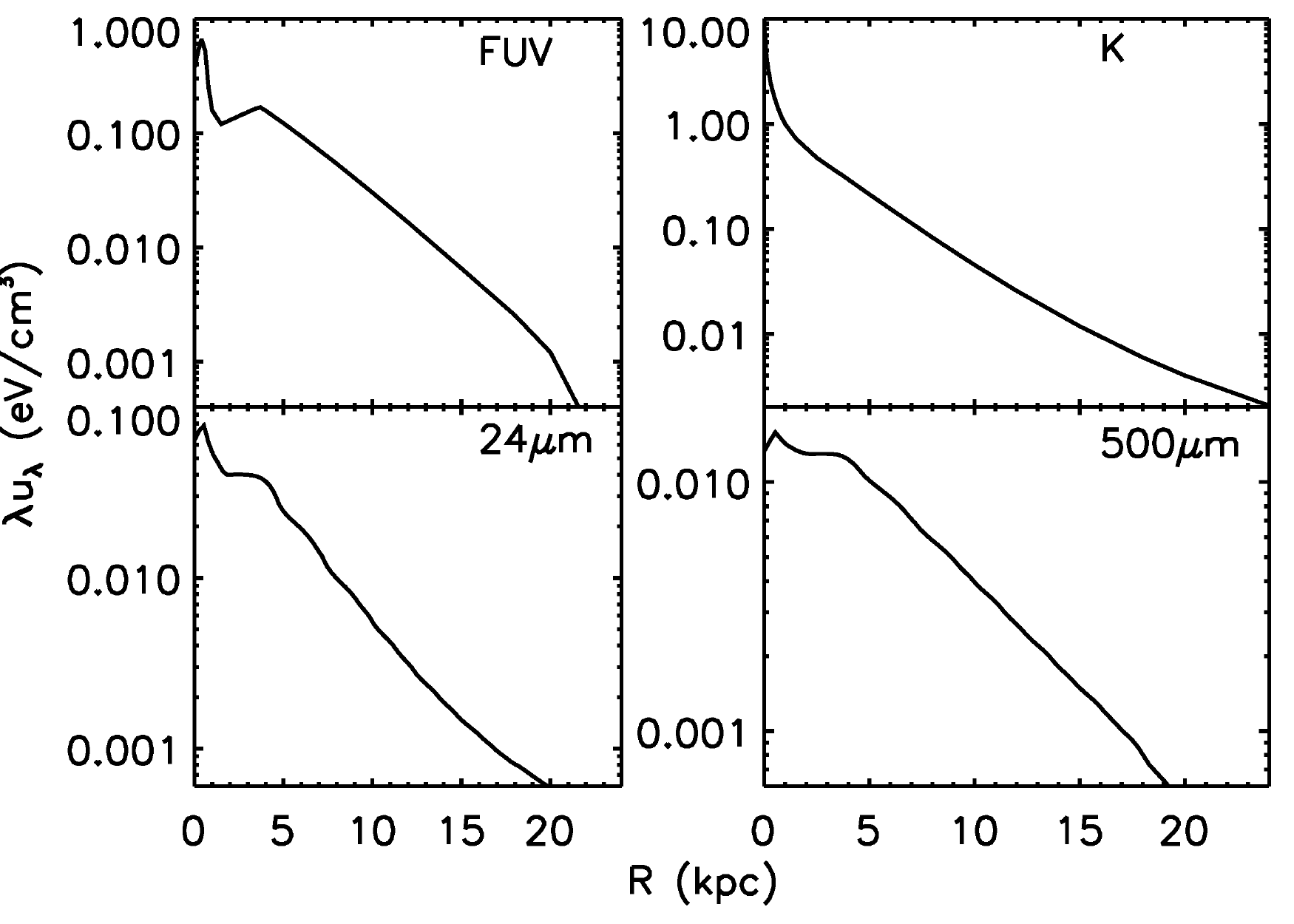}
\caption{Radial profiles of radiation fields for NGC~628, at selected wavelengths.}
  \label{fig:urad}
\end{figure}

The radial profiles of the energy density of the radiation fields of NGC~628 at selected wavelengths are shown by in Fig.~\ref{fig:urad}. The radiation fields are calculated in stellar light and dust emission. The variation with wavelength reflects the relative importance of the radiation fields from the four stellar components in the model. In the NUV band, the radiation fields of the inner thin stellar disk dominate the centre, attaining a maximum at about 500~pc. In the K band, the radiation fields from this component are swamped by those of the bulge and old stellar disk.
The radial profiles in dust emission are qualitatively similar across the sampled wavelengths. From short to long wavelengths, the main changes are a decline in the relative contribution from the inner thin dust disk and a shallowing slope from the outer disk.

In addition to the contribution of the different stellar components, the variation with wavelength of the radiation fields is also influenced by opacity effects. Thus, outwards of the inner radius, the slope of the radial profiles follows the trends for exponential disks, as described in Popescu \& Tuffs (2013). The slope is shallower in the FUV than in the K band, as expected for a more optical thick solution in the FUV than in the K-band. The slope in the K-band tends to a $R^{-2}$ decline, typical for optically thin solutions.

The opacity of the solution also influences the shape of the profile in the outer disk. Thus, in the FUV the radiations fields exhibit a truncation radius at around 20\,kpc, following the stellar truncation radius. The prominent break is due to the optical thick character of the solution, following the emissivity distribution. In the K band, where the solution is optically thin, the radiation fields do not show any sign of truncation, but instead follow the monotonic decline tending to a $R^{-2}$ function.

\section{discussion}
\label{sec:discussion}

Having modelled NGC~628 with a radiative transfer method that  self-consistently takes into account both the direct stellar light and the dust emission, we are now in a position to discuss the properties of this galaxy based on fundamental intrinsic properties.. 
What does this imply for the assemble of stellar populations and dust in this galaxy? In the following we consider several aspects.

\subsection{The contribution of young and old stellar populations in heating the interstellar dust}

\begin{figure}
\includegraphics[width=0.4\textwidth,angle=-0]{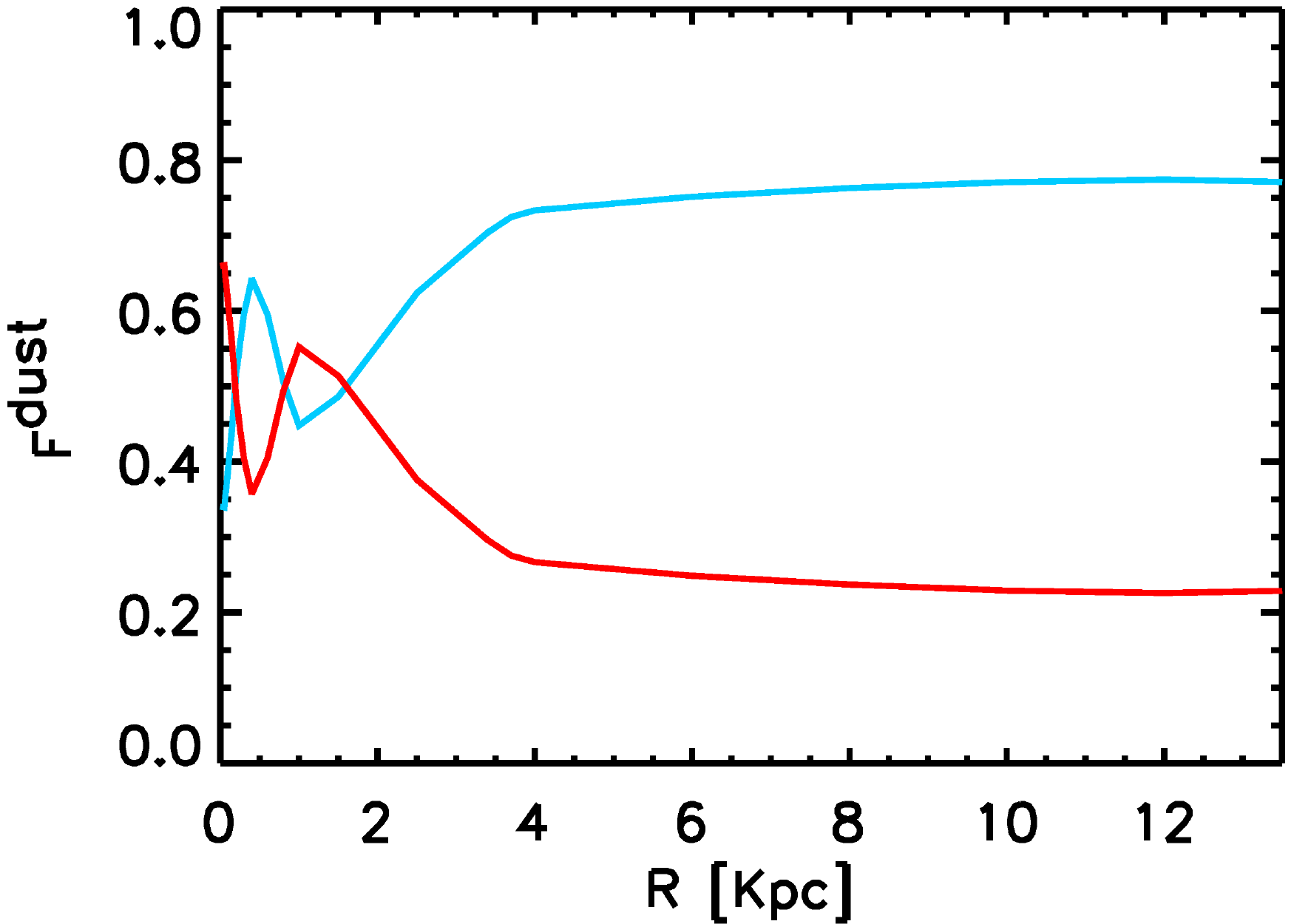}
\caption{Radial profiles of the fraction of stellar light from the young (blue) and old (red) stellar populations in heating the dust, $F^{\rm dust}_{\rm young}$ and $F^{\rm dust}_{\rm old}$.}
\label{fig:absorbed_fraction}
\end{figure}

In Sect.~\ref{subsec:global_sed} we found that  $71\%$ of the global dust emission is powered by the UV/optical photons emitted by stars in the thin stellar disk and inner thin stellar disk. Thus, most of the dust is heated by the young stellar populations. This also remains true on local scales, throughout most of the spatial extent of NGC~628.   Fig.~\ref{fig:absorbed_fraction} shows the spatial variation of the fraction of the stellar light from the young and old stellar populations, $F^{\rm dust}_{\rm young}$ and  $F^{\rm dust}_{\rm old}$, in the form of an azimuthally averaged radial profile. One can see that between 4 to 13 kpc, $F^{\rm dust}_{\rm young}$ is quite high, above $75\%$. At radii smaller than 4\,kpc this fraction decreases, with some variation, due to the contribution from other morphological components, like the inner thin disk and bulge. Thus, there is a peak in the  $F^{\rm dust}_{\rm young}$ at around 0.5 kpc, which is attributed to the inner thin stellar disk. The old stellar populations from the bulge dominate the dust heating in the very centre, with $F^{\rm dust}_{\rm old}\sim0.65$. 

The global value of 
$F^{\rm dust}_{\rm young}=0.71$ is the same as that of the Milky Way (Natale et al. 2021). Interestingly, the radial variation of this factor is also similar to that of our Galaxy, albeit with more variation in NGC~628. In both galaxies $F^{\rm dust}_{\rm young}$ is more or less constant, or slightly increasing outwards of the inner radius of the main disk. Furthermore, in both galaxies $F^{\rm dust}_{\rm young}$ decreases towards the centre, with old stellar populations from the bulge dominating the central heating. The only difference is that the decrease is monotonic in the Milky Way, while in NGC~628 there are oscillations, due to the presence of the inner thin disk.

\begin{figure}
  \includegraphics[width=1.0\linewidth]{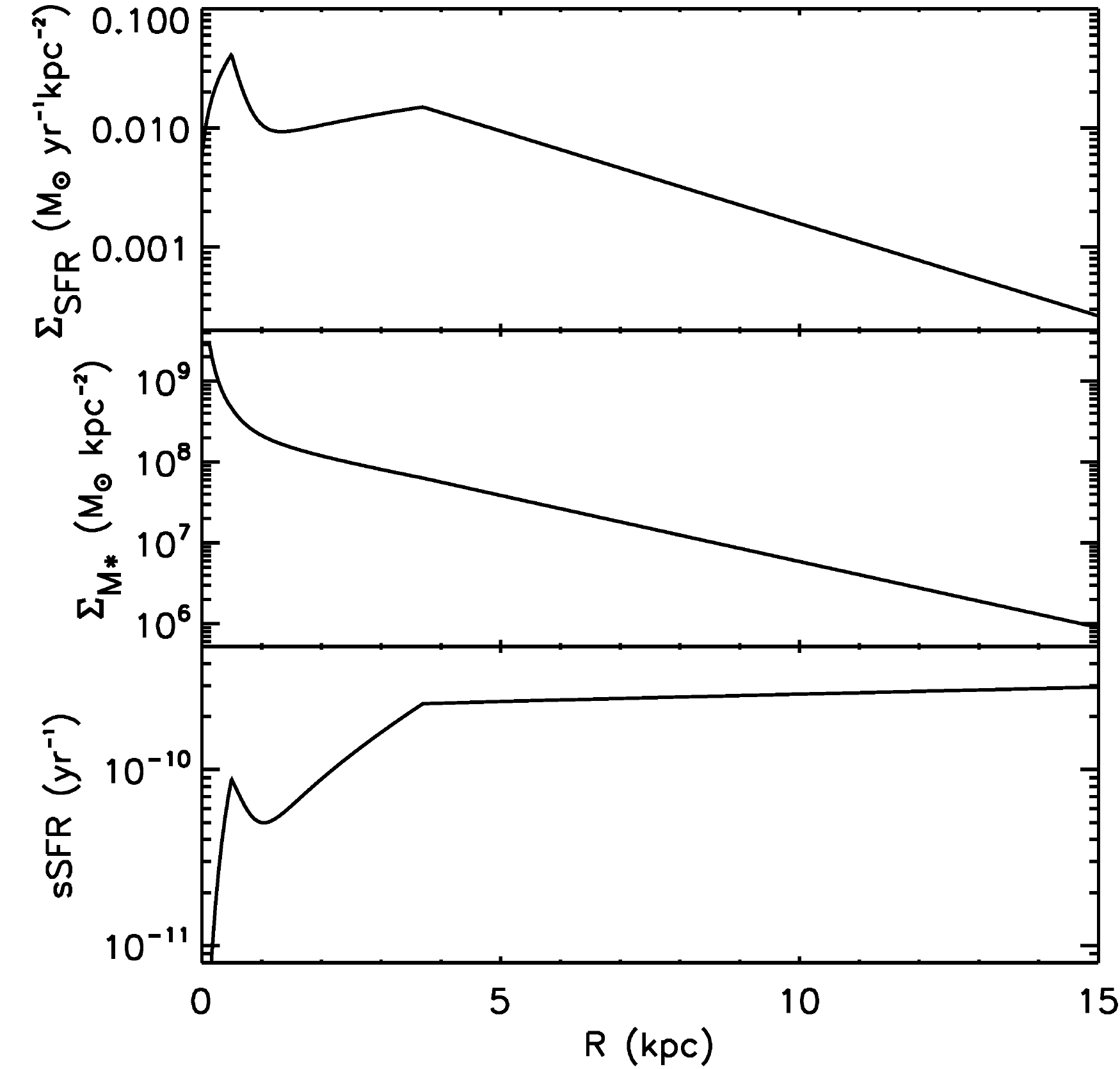}
\caption{Radial profiles of the SFR surface density, $\Sigma_{\rm SFR}$ (top), stellar mass surface density, $\Sigma_{\rm M_\star}$ (middle), and specific star formation rate (sSFR; bottom).}
  \label{fig:sfr_mass_ssfr}
\end{figure}

\begin{figure}
  \includegraphics[width=1.0\linewidth]{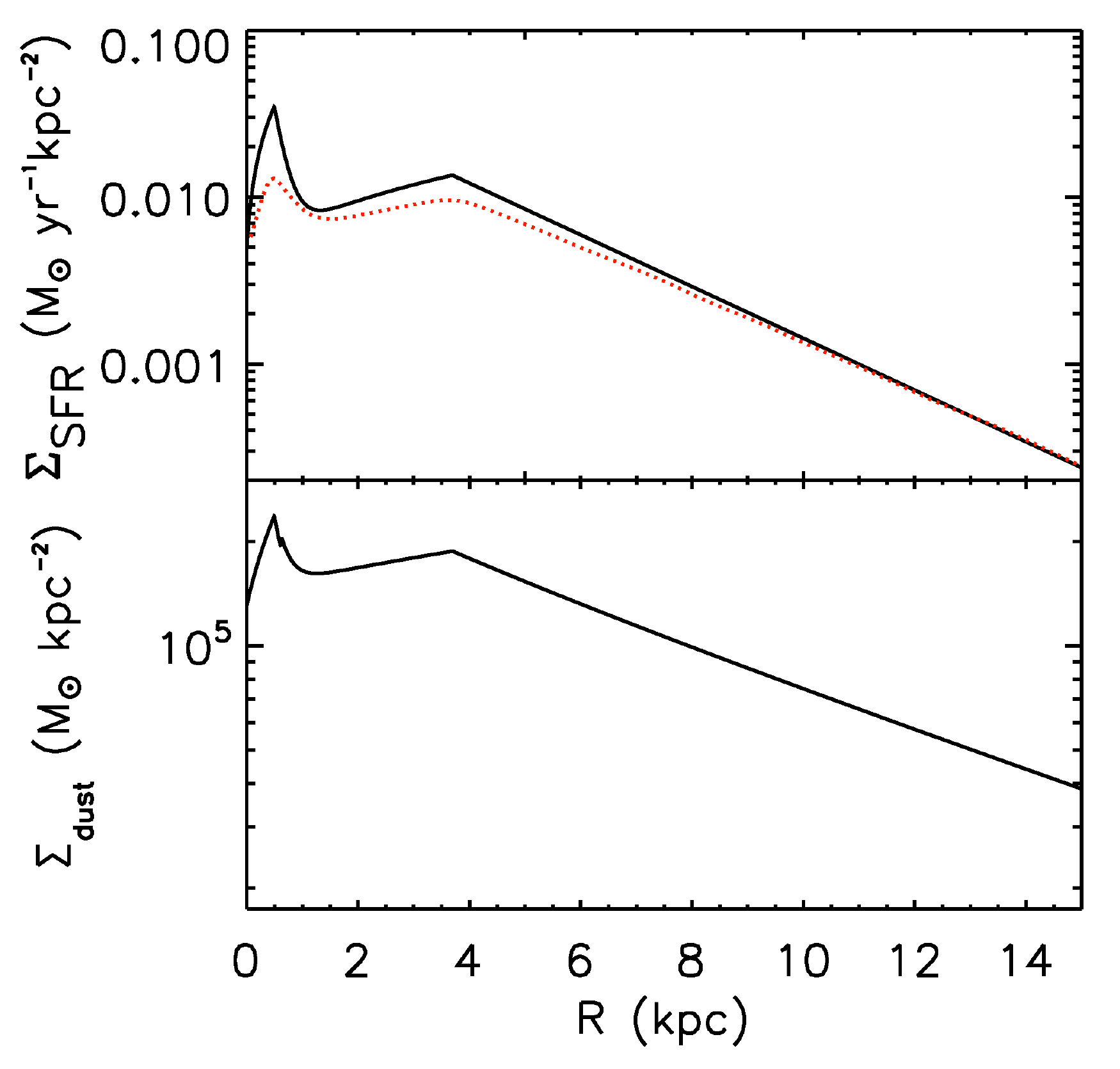}
\caption{Top: Radial profile of the SFR surface density, $\Sigma_{\rm SFR}$. The red dotted line is the $\Sigma_{\rm SFR}$ derived using the empirical relation from Bigiel et al. (2008). Bottom: radial profile of dust mass, $\Sigma_{\rm dust}$. }
  \label{fig:ssfr_bigiel}
\end{figure}

\subsection{The spatial distribution of SFR, $M_*$,  sSFR and $M_{\rm dust}$}
\label{sec:sfr_distrib}

The radiative transfer model allows us to derive not only integrated quantities, but even more important, spatial distributions, since the latter are more difficult to derive with empirical methods. In particular the radial profiles of surface densities of SFR, $M_*$, and $M_{\rm dust}$, and the radial profile of sSFR, are of prime importance, and are discussed below.

The surface density of SFR, $\Sigma_{\rm SFR}$, shows a pronounced peak at 
0.5\,kpc, at the inner radius of the inner thin disk, $R_{\rm in}^{\rm i-tdisk}$ (see upper panel in Fig.~\ref{fig:sfr_mass_ssfr}). The peak is obviously due to the contribution of the inner thin stellar disk. 
Beyond the inner radius of the thin stellar disk (3.7\,kpc), there is an exponential decrease in $\Sigma_{\rm SFR}$, following the exponential decrease of the thin stellar disk emissivity. The scale-length of this decrease is $h_{ \Sigma_{\rm SFR}}=(2.8\pm 0.2)$\,kpc, the same as the scale-length of the thin stellar disk, $h_{\rm s}^{\rm tdisk}$. 

The radial profile of $\Sigma_{\rm SFR}$ in NGC~628 was also derived in Casasola et al. (2017), using the empirical relation between GALEX-FUV and the WISE 22\,${\mu}$m from Bigiel et al. (2008):
\begin{equation}
 \Sigma_{\rm SFR}=3.2\times10^{-3}\times I_{22} + 8.1\times 10^{-2} \times I_{\rm FUV}
 \label{eq:bigiel}
\end{equation}
where $\Sigma_{\rm SFR}$ is in units of ${\rm M}_{\odot}{\rm yr}^{-1}{\rm kpc}^{-2}$, and $I_{\rm 22}$ and $I_{\rm FUV}$ are the 22\,$\mu$m and FUV intensities, respectively, in units of MJy\,sr$^{−1}$. Casasola et al. (2017) also fitted an exponential between 3.1 and 11.5 kpc (values adjusted to our adopted distance of NGC~628), and found a scale-length (also adjusted to our adopted distance) of $h_{{\Sigma}_{\rm SFR}}=(3.47\pm 0.12)$\,kpc. Their derived value for $h_{{\Sigma}_{\rm SFR}}$ is much larger than our value. 
The first reason of discrepancy is clearly the different range in radial values adopted for the fit. 
A second and perhaps more subtle reason is the shape of the derived profile. To understand this second point, we derived ourselves the profile of ${\Sigma}_{\rm SFR}$ using the empirical relation from Bigiel, from Eq.~\ref{eq:bigiel}, applied to our fitted model profiles. The same results are obtained if the relation is directly applied to data, since our model represents a very good fit to the observed profiles. The result is plotted with red dotted line in the upper panel of Fig.~\ref{fig:ssfr_bigiel}, for comparison with our model profile of $\Sigma_{\rm SFR}$ (black line). It is clear that there is a marked difference between the two profiles. In particular in the inner thin disk region, but also at the inner radius of the disk, the empirical relation of Bigiel provides a much lower ${\Sigma}_{\rm SFR}$. Overall the empirically derived profile is more featureless than our model profile. It also has a shallower outer slope, leading to the larger derived scale-length. To conclude, the profile of ${\Sigma}_{\rm SFR}$ derived with our RT model is quite different from that produced with the Bigiel empirical relation.

The surface density of stellar mass, $\Sigma_{\rm M*}$, was derived using the 3.6\,${\mu}$m stellar luminosity profile as a proxi for the stellar mass, calibrated to the global stellar mass of $1.21\times 10^{10}\,{\rm M}_{\odot}$.
The profile (middle panel in Fig.~\ref{fig:sfr_mass_ssfr}) 
has a pronounced peak in the centre, mostly due to the contribution of the bulge. Beyond the bulge region, the radial profile of $\Sigma_{\rm M*}$ starts to decrease, followed by an exponential decrease, with a scale-length $h_{\Sigma_{\rm M*}}=2.7\pm0.2$\,kpc (by definition,  the same as the $h_{\rm s}^{\rm disk}$ at 3.6\,$\mu$m). The main difference between the profiles of $\Sigma_{\rm SFR}$ and $\Sigma_{\rm M*}$ is that the peak is not at the same radial position. There is a 0.5\,kpc shift between them, corresponding to the spatial shift between the inner thin disk and the bulge, the main  contributors to the corresponding peaks.

As in the case of $\Sigma_{\rm SFR}$, the radial profile of $\Sigma_{\rm M*}$ can be compared with that derived in Casasola et al. (2017). Their fitted values are $h_{\Sigma_{\rm M*}}=2.92\pm0.15$\,kpc (adjusted) when using IRAC data as a tracer of stellar mass, or $h_{\Sigma_{\rm M*}}=2.54\pm0.15$\,kpc (adjusted) when using SDSS data as a tracer. Since we also used IRAC data to trace stellar mass, we consider the former value for the comparison. This would indicate that Casasola et al. derive a slightly larger value for $h_{\Sigma_{\rm M*}}$ than us, although almost consistent within errors. This is perhaps not surprising, since the NIR profiles of NGC~628 can be considered free from dust effects, and as such no differences due to a different treatment of attenuation effects should occur.

The radial profile of  sSFR (see bottom panel of Fig.~\ref{fig:sfr_mass_ssfr}) is derived by dividing the profile of SFR to that of ${\rm M*}$. The main feature of the profile is an almost constant value between the inner radius of the disk (3.7\,kpc) and the outer truncation radius of the disk (20\,kpc). At radii smaller than 3.7\,kpc the sSFR starts to decrease, but not monotonically. It has an oscillatory behaviour, with a secondary peak at the position of the inner thin disk. In the centre the sSFR is severely depressed. 

The fitted scale-length of $\Sigma_{\rm SFR}$ and $\Sigma_{\rm M*}$ show that the SFR has a slightly shallower decrease with radius than the stellar mass. This result is a direct consequence of the fact that the SFR traces the UV emitting disk, while the stellar mass traces the old stellar disk, and that there is not a marked difference between the old and young stellar disk scale-lengths in NGC~628. This result is different from that obtained in Casasola et al. (2017), and the reasoning behind the discrepancy was explained above. Other galaxies were found to have a more pronounced difference in the slope of SFR and M* profiles, due to a more pronounced difference between the scale-length of UV and NIR emitting disks (Popescu et al. 2000, Natale et al. 2021). Even M33 has a more pronounced difference in scale-lengths (Thirlwall et al. 2020) than NGC~628. A decrease in the scale-length of the stellar emissivity with increasing wavelength has been attributed to the fact that disks of spiral galaxies grow from inside out, as predicted by semi-analytical hierarchical models for galaxy formation (Mo et al. 1998). One would predict the stellar populations to be younger and have lower metallicity in the outer disk than in the inner disk, such that local Universe galaxies should be intrinsically larger at the shorter wavelengths where light from the young stellar population is more prominent. The milder difference between the UV and NIR emitting disks of NGC~628 may indicate that an inside out scenario may not have effectively operated in this case.

The radial profile of dust mass surface density, $\Sigma_{\rm M_d}$, is shown in the bottom panel of Fig.~\ref{fig:ssfr_bigiel}. The profile has a peak at the position of the inner thin disk, and a secondary peak at the inner radius of the dust disk. At larger radii the dust mass follows the exponential decrease of the dust disk, with a scale-length of $h_{\Sigma_{\rm M_{\rm d}}}=7.1\pm 0.5$\,kpc, which is almost the same (as expected) as the scale-length of the dust disk, $h_{\rm d}^{\rm disk}$. This shows that the dust mass has a rather flat radial distribution with respect to that of SFR and M*. This is not surprising, since the dust distribution is known to have a larger scale-length than either the distribution of older stars (Xilouris et al. 1997, 1998, 1999, de Geyter et al. 2014) or that of young stars (Popescu et al. 2000, Popescu et al. 2011). In addition, the dust distribution has been also found to be more radially extended than the optical emission, in particular following extended HI disks (Popescu \& Tuffs 2003). 

The result obtained for $h_{\Sigma_{\rm M_{\rm d}}}$ can be compared with that from Casasola et al. (2017). Their value for NGC~628 is $h_{\Sigma_{\rm M_{\rm d}}}=6.16\pm0.25$\,kpc (adjusted to our distance estimate), which is somewhat smaller than our derived value. The difference is probably due to the dust mass being derived locally, for each radial bin, in Casasola et al. (2017), while we use a radiative transfer method. 

The analysis above showed that the scale-length of the dust mass surface density is a factor 2.6 larger than that of the SFR or $M_*$. These factors are larger than those found in Casasola et al. (2017), with the difference most likely originating in the different methodology used: empirical in Casasola et al. and RT methods in this paper. Overall, the large factors found for NGC~628 are above the average. They are larger than the factors derived for the Milky Way in Natale et al. (2021). Even in Casasola et al. (2017), the factors they derived, although smaller than in our analysis, are still larger than the average of their sample. This indicates a more radially extended distribution of dust in NGC~628 than on average. Perhaps there is no coincidence that NGC~628 has an extended HI disk. The large amounts of dust at larger radii may indicate some efficient mechanism of either transporting grains from the denser,  inner regions of the disk to the outer regions, or some external origin for the dust. In Popescu \& Tuffs (2003) two possible scenarios were introduced to account for the dust associated with the extended disk of NGC~891, one in which grains could be transported via the halo to larger radii, and one in which dust grains and gas could be transported via diffusion triggered by macro turbulence. These 
scenarios may also explain the distribution of dust in NGC~628. An external origin could invoke some accretion from the intergroup medium of NGC~628, where the accreting material would already contain dust grains (see Natale et al. 2010).  This scenario is less likely, taken into account our findings that an inside-out disk growth through accretion has not played a major role in NGC~628.
However, a past interaction with some of its two dwarf satellites may be a possibility. For example the extended HI disk of NGC~891 and its associated dust has been associated to non-primodial material from the inner 
disk, that has been perturbed by a nearby companion (Bianchi \& Xilouris 2011). Again, this situation may also apply to NGC~628.

\subsection{The attenuation curve of NGC~628}

\begin{figure}
\includegraphics[width=0.4\textwidth,angle=-0]{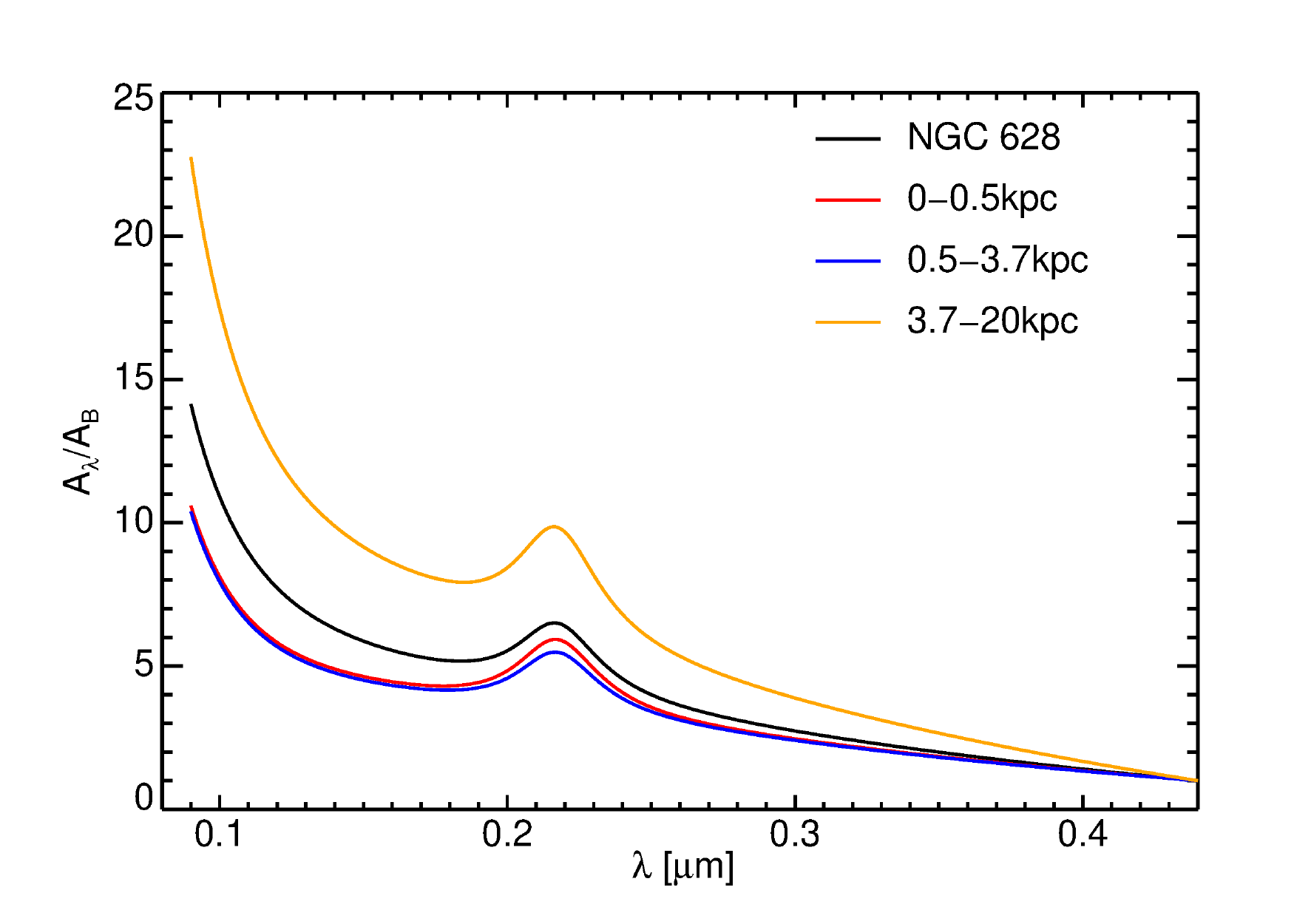}
\caption{The attenuation curve of NGC~628 (at its observed inclination), depicted with black line. The coloured lines show the attenuation curve within different radial bins, chosen to coincide with the location where the bulk of the emission from  the bulge, inner thin disk and disk occurs.}
\label{fig:attencurve}
\end{figure}

The attenuation curve of NGC~628 was produced by taking the ratio between the apparent (dust-attenuated) and intrinsic (dust-deattenuated) model maps, at the various sampled wavelengths. The derived values at the sampled points were then fitted with the functional form from Salim et al. (2018) (see their Eqs. 8 and 9). In Fig.~\ref{fig:attencurve} we show with black line the resulting curve, integrated over the whole extent of the galaxy, and normalised to the B-band attenuation. 
The coefficients of the fit are listed in Table~\ref{tab:fit_coeff}. 
We also divide the galaxy in 3 radial bins (0-0.5; 0.5-3.7; 3.7-20\,kpc), and plot the corresponding attenuations curves in red, blue and orange (with fitting coefficients also listed in Table~\ref{tab:fit_coeff}. In the division of the bins we took the inner radius of the inner thin disk and disk into account.  

The curve of the galaxy (as a whole) is steeper than that of the inner region, but shallower than that of the outer regions. This trend is similar to that found for M33 in Thirlwall et al. (2020), and can be explained by opacity effects: more optically thin would mean more variation between shorter and longer wavelengths, producing a steeper attenuation curve.

\begin{table}
\caption{Coefficients for the best-fitting function to the attenuation curve of NGC~628.}
  \begin{tabular}{ | l | cccccc }
 \multicolumn{4}{c}{}    \\
 \hline
 & a(0) & a(1) & a(2) & a(3) & B & $R_{\rm B}$ \\
\hline
global &-4.02 & 2.10 & -0.28 & 0.0168 & 1.50 & 1.010 \\
0.0-0.5\,kpc & -3.87 & 2.10 & -0.30 & 0.0169 & 1.68 & 1.073 \\
0.5-3.7\,kpc & -3.77 & 2.10 & -0.30 & 0.0170 & 1.47 & 1.000 \\
3.7-20.\,kpc & -4.44 & 2.22 & -0.30 & 0.0176 & 1.47 & 1.095 \\
\hline
 \end{tabular}
 \label{tab:fit_coeff}
 \end{table}

\begin{figure*}
\includegraphics[width=0.6\textwidth,angle=-0]{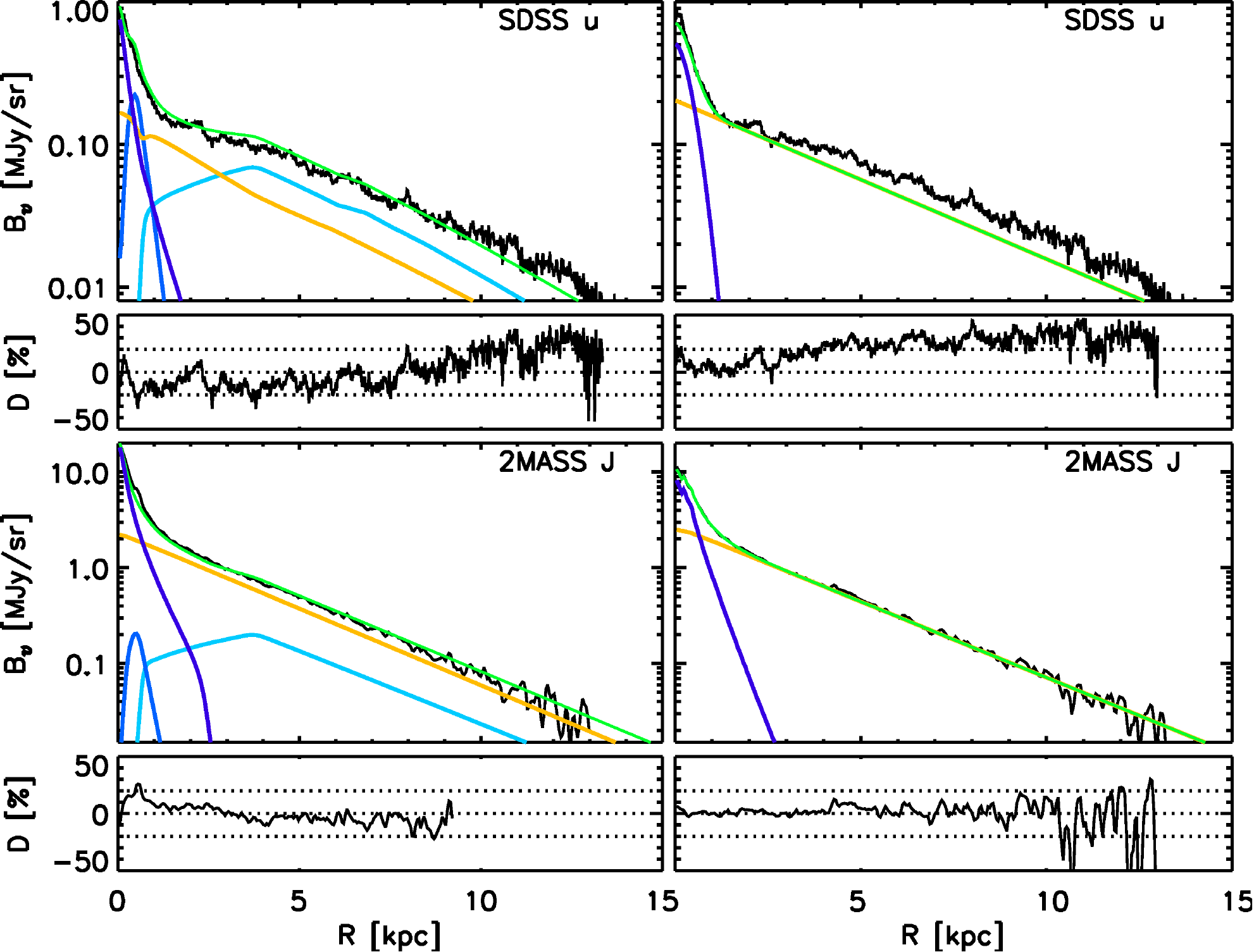}
\caption{Comparison between the model surface brightness profiles obtained from our RT analysis using multicomponent fits (left) and the corresponding ones obtained from GALFIT using B/D decomposition (right). The comparison is done at the wavelengths corresponding to the SDSS u band and the 2MASS J band.}
\label{fig:uj_galfit}
\end{figure*}

\subsection{Multicomponent fits and projection effects on the surface-brightness photometry}

As mentioned in  the description of our optimization procedure in Sect.~\ref{sec:optim}, we initially performed a standard bulge-disk decomposition on the u\,g\,r\,i\,z\,J\,H\,K observed images utilised in our analysis, using the GALFIT package (Peng et al. 2002). This allowed us to have a first guess estimate of the initial parameters for the scale-length of the stellar disk $h_{\rm s}^{\rm disk}$, the Sersic index of the bulge $n_{\rm s}$, and the effective radius of the 
bulge $R_{\rm eff}$. As expected, the final results obtained from our modelling approach differ significantly from these initial guesses, due to several reasons. Firstly, our RT analysis considers more than only 2 structural components. 
 Secondly, GALFIT only deals with apparent quantities, which do not include  effects due to the  projection of 3D quantities and to dust. In the following we shall discuss both the effect of multicomponent fits and the projection effects.

\begin{table}
\caption{Photometric parameters derived from the bulge-disk decomposition with GALFIT: apparent bulge Sersic index $n_{\rm s, app}$, apparent bulge effective radius $R_{\rm eff}$ and apparent disk scale-length $h_{\rm s}$. For comparison the intrinsic scale-length $h_{\rm s}$ derived from the RT modelling is also listed. }
  \begin{tabular}{ | l | cccc }
 \multicolumn{3}{c}{}    \\
 \hline
band & $n_{\rm s,app}$ & $R_{\rm eff, app}$ [pc] & $h_{\rm s, app}$  [pc] & $h_{\rm s}^{\rm disk}$ [pc] \\
\hline
u & $0.91\pm0.05$ & $427\pm10$ & $3909\pm16$ & 3000\\
g & $1.37\pm0.01$ & $544\pm4$ & $3534\pm2$   & 2900\\
r &  $1.64\pm0.01$ & $659\pm4$ & $3404\pm3$  & 2900\\
i & $1.47\pm0.01$ & $623\pm5$ & $2955\pm8$   & 2800\\
z & $1.58\pm0.03$ & $673\pm11$ & $3155\pm7$  & 2800\\
J & $1.29\pm0.02$ & $567\pm6$ & $2736\pm6$   & 2700\\
H & $1.36\pm0.01$ & $598\pm6$ & $2695\pm8$   & 2700\\
K & $1.26\pm0.02$ & $526\pm7$ & $2506\pm11$  & 2700\\
\hline
 \end{tabular}
 \label{tab:galfit}
 \end{table}
 
In Fig.~\ref{fig:uj_galfit} we show a comparison between our model fits (left) and the GALFIT model (right) for the  u and J bands used in the initial B/D decomposition
(comparison for all ugrizJHK bands are shown in the Appendix, in Figs.~\ref{fig:galfit_SDSS}, \ref{fig:galfit_SDSSi}, and \ref{fig:galfit_2MASS}). In the following we shall refer to the parameters derived from the GALFIT bulge-to-disk decomposition as \lq\lq apparent\rq\rq, and we shall use the subscript \lq\lq app\rq\rq\ for them. They are listed in Table~\ref{tab:galfit}: 
the apparent Sersic index of the bulge $n_{\rm s,app}$, the apparent effective radius of the bulge $R_{\rm eff, app}$, the apparent scale-length of the disk $h_{\rm s, app}$ together with our derived $h_{\rm s}$ for ease of comparison. We remind the reader that in our model the intrinsic Sersic index $n_{\rm s}=2$ and $R_{\rm eff}=820$\,pc (see Table~\ref{tab:param}).

Inspection of the profiles in the u band (Fig.~\ref{fig:uj_galfit}) clearly shows that the main discrepancy between the two type of model fits is due to the different number of structural components used. In addition to the disk and bulge, our model contains a inner thin stellar disk and a thin stellar disk harboring young stellar populations. Their existence was inferred through the panchromatic analysis, in particular from the UV data in combination with the dust emission maps. The initial GALFIT decomposition using only the optical/NIR data was insensitive to the existence of these extra components. As expected from young components, the contribution of the inner thin stellar disk and stellar disk decreases with increasing wavelenth, with a monotonic decrease from the u to the z band, becoming negligible in the JHK bands (see Figs.~\ref{fig:galfit_SDSS}, \ref{fig:galfit_SDSSi}, \ref{fig:galfit_2MASS}). Because of this, the GALFIT parameters are more influenced by this effect at shorter wavelength. 
Thus, in the u band, where the discrepency is largest, the GALFIT does not even produce a satisfactory fit to the data.   It is clear that, in the absence of a designated component to describe the inner thin disk, this  is, by necessity, incorporated in the bulge in GALFIT (see Fig.~\ref{fig:uj_galfit}), with the consequence that the bulge emissivity is fitted with a shallower slope and a smaller effective radii. The GALFIT parameters are $n_{\rm s,app}=0.91$ and $R_{\rm eff}=427$\,pc (see Table~\ref{tab:galfit}), much smaller than our values. Similarly, the thin stellar disk is incorporated in the disk in GALFIT, resulting in a larger scale-length for the disk ($h_{\rm s,app}=3909$ pc as opposed to $h_{\rm s}=3000$\,pc), and an underestimation of the emission outside of the bulge region. As expected, from u to z band, $n_{\rm s,app}$ increases, $R_{\rm eff,app}$ decreases and $h_{\rm s,app}$ decreases (see Table~\ref{tab:galfit}), getting closer to our derived values, as the contributions of the inner thin stellar disk and thin stellar disk become smaller.

In the NIR (JHK) the young stellar components become negligible in the inner disk, and the main cause of difference in parameter values is due to projection effects, since dust effects are negligible at these wavelengths. As explained in Pastrav et al. (2013a), projection effects arise even in the absence of dust, causing the fitted functions used in surface-brightness photometry codes to imperfectly recover the structure of real disks and bulges due to the fact that these functions describe infinitely thin templates, in contrast to real disks and bulges, which have a thickness. Thus, the additional vertical distribution of stars superimposed on the radial distribution produces isophotal shapes which differ from those predicted by an infinitely thin template. 
 Of course these artifacts are specific to routines that perform bulge-disk 
decomposition using  analytical infinitely thin dustless templates, like the GALFIT, but this is the only feasible approach unless a full RT modelling is undertaken. 

To quantify projection effects on the derived Sersic index we use the formalism from Pastrav et al. (2013a) and look at the difference between the apparent Sersic index $n_{\rm s,app}$ derived in the bulge-disk decomposition in the JHK bands (which is around 1.3-1.4) and  the derived intrinsic Sersic index of our RT analysis (which is 2):
\begin{align}
corr^{{\rm proj}}(n_{\rm s})=n_{\rm s, app}-n_{\rm s}
\end{align}
We derive $corr^{{\rm proj}}(n_{\rm s})=-0.6$. Pastrav et al. derived 
  $corr^{{\rm proj}}(n_{\rm s})=-0.4$, independent on inclination and wavelength (see their Fig.~5)\footnote{The projection and the dust effects have been quantified  in Pastrav et al. (2013a,b). The quantification of these effects is given in terms of corrections between apparent and intrinsic quantities, (as available at \newline http://cdsarc.u-strasbg.fr/viz-bin/qcat?J/A+A/553/A80 and \newline 
http://cdsarc.u-strasbg.fr/viz-bin/qcat?J/A+A/557/A137)  using the same RT analysis as in this paper, but on simulations of ideal exponential disks and Sersic bulges.}. Taken into account the more complex geometry of NGC628 (as compared to the disk-bulge geometry from the simulations of Pastrav et al.), it is reasonable to expect a slightly different, but not far off, value for the projection effects. To conclude, the Sersic index close to 1 derived by GALFIT is the index of the 2 dimensional Sersic distribution of the projected (decomposed) image of the  bulge, while the Sersic index of 2 derived by our RT code is the index of the 3-dimensional distribution that in projection would produce the (decomposed) image of the bulge.

The $R_{\rm eff}$ is correlated with the value of $n_{\rm s}$, with $R_{\rm eff}$ decreasing as $n_{\rm s}$ decreases. As such, it is to be expected that the GALFIT derived $R_{\rm eff}$  is smaller than our derived $R_{\rm eff}$.

\subsection{Amplification of stellar light}

The effects of dust in attenuating the optical/NIR stellar light in NGC628 are rather small. We can safely consider that in the K band and longer wavelengths the overall dust attenuation is zero. At shorter optical wavelengths though we predict a small level of amplification due to scattering. This is due to the fact that the dust opacity is very low, but not zero. It is in this regime that the scattering dominates over absorption, producing a small level of amplification in the face-on view, which is the case of NGC628. Since the effects are small, RT scattering calculations with increased sampling along the rays have been performed, to test that no computational artifacts affect the results. We found constant values with increasing sampling, proving that negative attenuation is real within the remit of our model.

We find that negative attenuation is more pronounced for the disk (see Table~\ref{tab:amplification}), with the maximum amplification in the surface brightness distribution in the H band, with values  reaching up to -0.9 mag/pixel. For the integrated emission in the disk, the maximum amplification was found in the J-band, of -0.05 mag. But even in the g band there is some amplification, but only in the outer disk, of -0.1 mag/pixel.

The thin disk is predicted to have less amplification (see Table~\ref{tab:amplification}), reaching a maximum of -0.5 mag/pixel for the surface brightness distribution in the H-band. For the integrated flux of the thin disk the maximum negative attenuation is -0.06 mag in the J and z band.
As expected, we do not find any amplification in the inner thin disk and bulge. 

The level of negative attenuation derived for NGC628 is similar with the results obtained in our generic models from Tuffs et al. (2004). We need however to emphasise that  amplification is predicted for diffuse distributions only. In NGC628 this would be the case for the inter-arm emission, in particular in the outer disk. We would obviously not expect amplification to happen at the locations of the spiral arms. 

Observational studies of  outer disks of spiral galaxies have shown that amplification due to dust scattering could explain some of the faint emission. Thus, Bland-Hawthorn et al. (2005) detected an extremely faint outer disk in the spiral galaxy NGC300, extended out to 10 scale-lengths, and attributed some of the emission to dust scattering. Faint emissions were also detected in the outer disks of NGC5383 by Barton \& Thompson (1997) and NGC4123 by Weiner et al. (2001), possible also having a component of scattered light.

 \begin{table}
 
\caption{The attenuation of the disk and thin disk in a few optical bands where amplification occurs. The table lists both maximum attenuation values of individual pixels of the maps, $a_{\rm disk}$  and  $a_{\rm tdisk}$, as well as total attenuation for the integrated flux densities, $A_{\rm disk}$  and  $A_{\rm tdisk}$.}
  \begin{tabular}{ l c c c c c}
    \hline
 $\lambda$ ($\mu$m ) & filter & $a_{\rm disk}$  &   $a_{\rm tdisk}$  & $A_{\rm disk}$      & $A_{\rm tdisk}$    \\
                     && mag/pix        & mag/pix               & mag               & mag     \\      
 \hline
 &&&&&\\
 0.48 &g&-0.10&-0.3 &   0.11 &   0.02  \\
 0.62 &r&-0.15 &-0.15 &   0.02 &  -0.1 \\
0.76 &i&-0.25 &-0.20 &  -0.01 &  -0.1  \\
0.91 &z&-0.5&-0.3&-0.04  & -0.06 \\
1.26 &J&-0.5&-0.4&-0.05  & -0.06    \\
1.66 &H& -0.9&-0.5&-0.03  & -0.05     \\
 \\
\hline
 
 \end{tabular}
 \label{tab:amplification}
 \end{table}

\begin{figure}
  \includegraphics[width=0.75\linewidth]{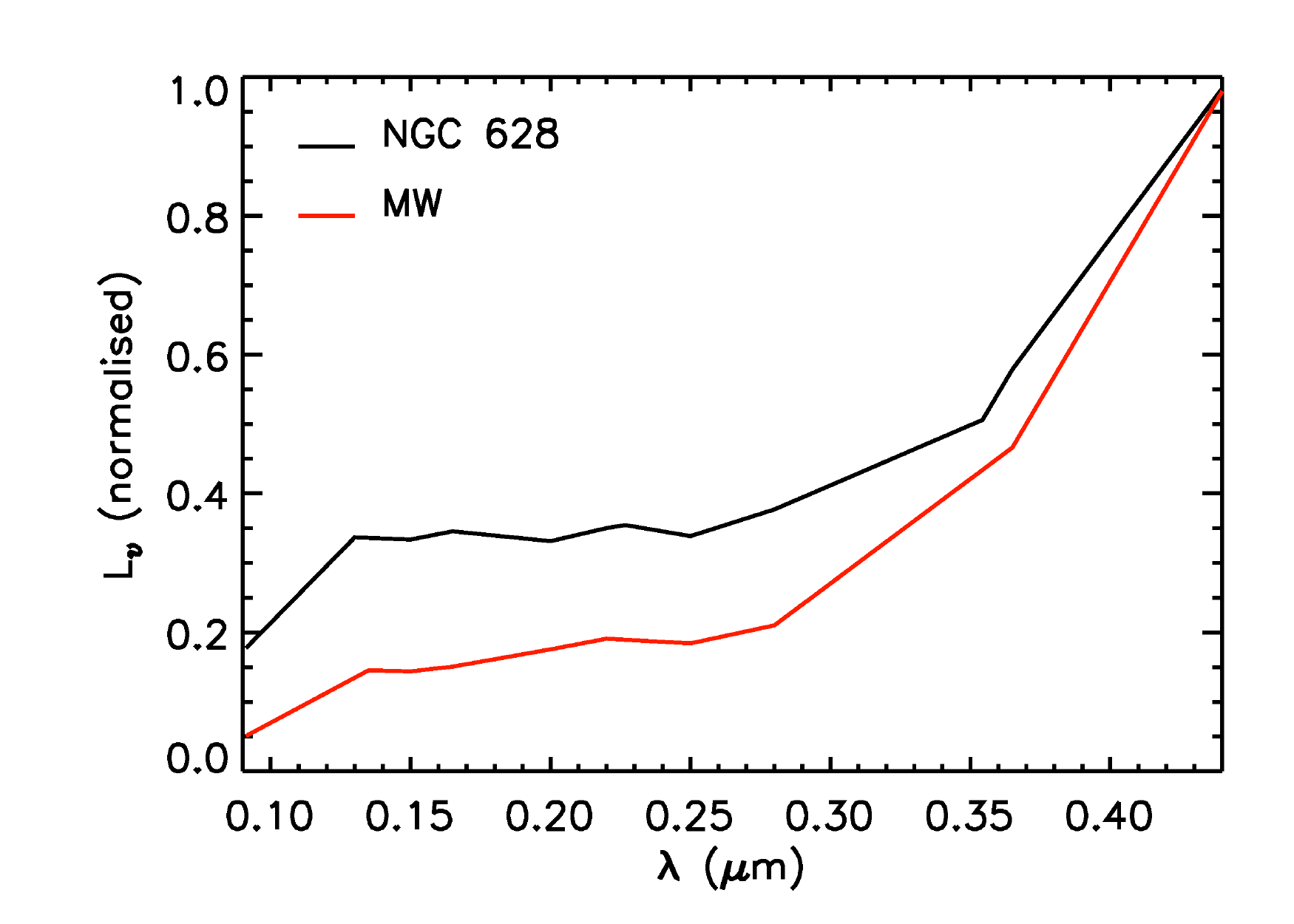}
\caption{UV-to-B band intrinsic SEDs of NGC 628 (black line) and the Milky Way (red line) normalised to their respective B band luminosities. The solution for the Milky Way is taken from Natale et al. (2021)}
  \label{fig:sed_n628_mw}
\end{figure}

\subsection{Comparison of NGC~628 with the Milky Way}

In the analysis presented in this paper we saw that some of the intrinsic parameters of NGC~628 are similar to those of the Milky Way. For comparison purposes we use the parameters of the Milky Way derived in Natale et al. (2021). Here we look in more details at this comparison. 

When comparing the global properties, we find some remarkable similarities. The fraction of stellar light from the young stellar population in heating the dust is the same for both NGC~628 and the MW, with $F_{\rm young}^{\rm dust}=0.71$. The fraction of the diffuse component to the total dust luminosity is also the same within the errors, with $0.79\pm 0.2$ and $0.81\pm 0.2$ for NGC~628 and the MW, respectively. The SFR surface density is also the same within errors, with $\Sigma_{\rm SFR}=(1.6\pm 0.2) 10^{-3}\,{\rm M}_{\odot}{\rm yr}^{-1}{\rm kpc}^{-2}$ for NGC~628 and $\Sigma_{\rm SFR}=(2.0\pm 0.3) 10^{-3}\,{\rm M}_{\odot}{\rm yr}^{-1}{\rm kpc}^{-2}$ for the MW. Both NGC~628 and the MW have the same face-on dust opacity at the inner radius of the disk, of $\tau_{\rm B}^{\rm f}(R_{\rm in}^{\rm disk})=1.47\pm 0.12$ and 
$\tau_{\rm B}^{\rm f}(R_{\rm in}^{\rm disk})=1.48\pm 0.10$, respectively. All these are consistent with NGC~628 being a quiescent spiral, with the dust luminosity powered by the young stellar population and being dominated by the diffuse component, and being moderately optically thick at the inner radius, in exactly the same proportions as the Milky Way. So from this point of view NGC~628 can be considered an analogue of our Galaxy. This is quite surprising, taking into account that the Milky Way 
has been classified as an Sb-Sbc (see review of Bland-Hawthorne \& Gerhard 2017) or possible as an even earlier type, while NGC~628 is an Sc spiral. The Milky Way has also a redder UV-optical SED than NGC~628, as revealed by Fig.~\ref{fig:sed_n628_mw}, where we plotted the intrinsic SED for the two galaxies. In addition, the Milky Way hosts a bar while NGC~628 does not. Otherwise, in the optical range the two galaxies have comparable sizes. They are both  dominant galaxies in a group.

\begin{figure}
  \includegraphics[width=1.0\linewidth]{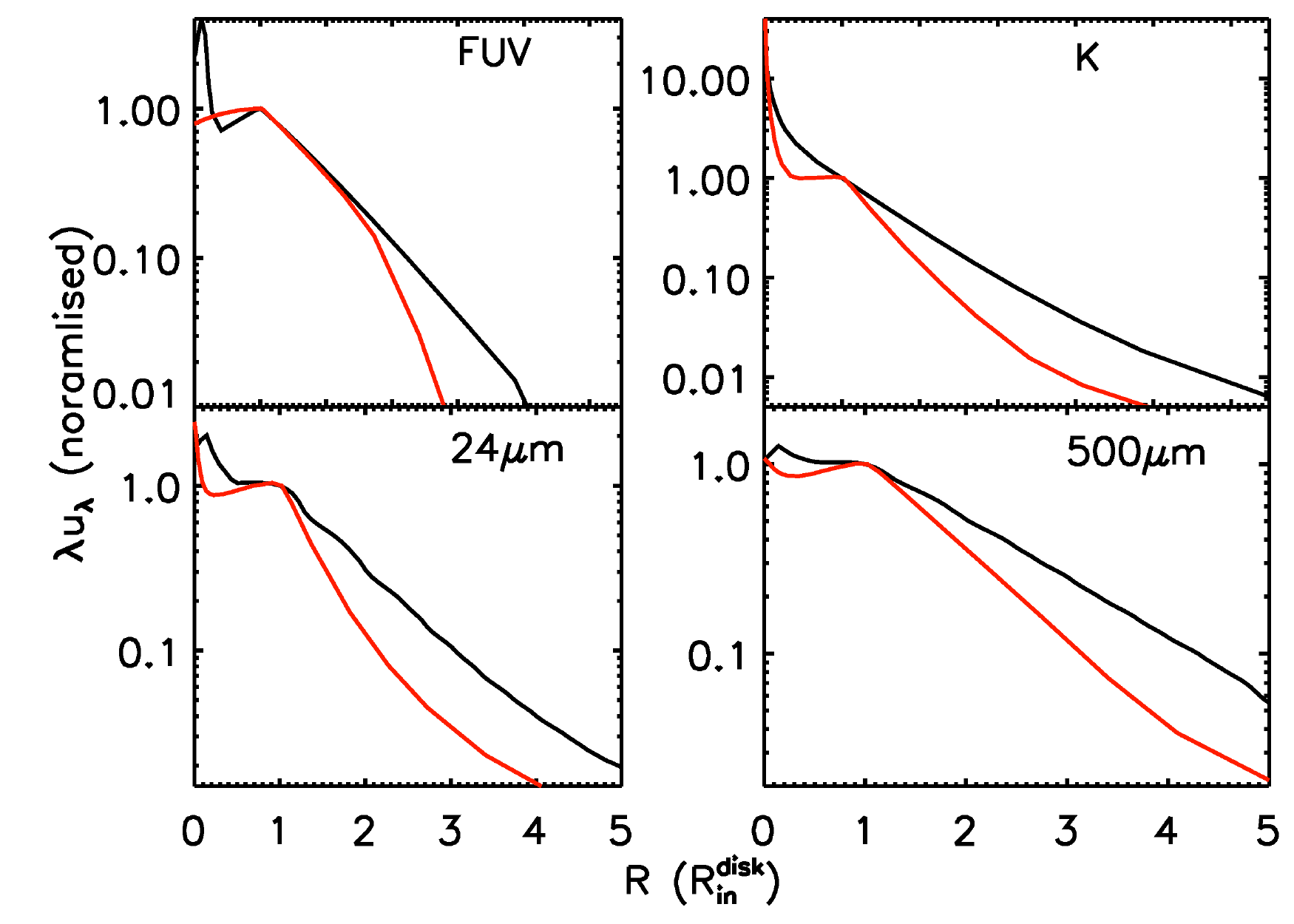}
\caption{Radial profiles of normalised radiation fields for NGC~628 (black line) and the Milky Way (red line). The radial distance is given in units of $R_{\rm in}^{\rm disk}$, and the energy densities of the radiation fields, $u_{\lambda}$, are normalised to their value at $R_{\rm in}^{\rm disk}$. The solution for the Milky Way is taken from Popescu et al. (2017).}
  \label{fig:urad_N628_mw}
\end{figure}

When comparing the spatial distributions though, differences can be found, which are easily understood in terms of differences in morphology. The distribution of radiation fields indicates the presence of the bar in the Milky Way, and the presence of the inner thin disk in NGC~628. This is illustrated in Fig.~\ref{fig:urad_N628_mw}, where we plot the normalised radial profiles of the radiation fields of the two galaxies. The most obvious comparison is by considering the radial distance in units of the inner radius of the disk, $R_{\rm in}^{\rm disk}$, and normalising the energy density to the value at $R_{\rm in}^{\rm disk}$. One can see that at radii smaller than $R_{\rm in}^{\rm disk}=1$, the profile of NGC~628 is shaped by the inner thin disk, while that of the MW is more flat, due to the contribution of the bar. For $R_{\rm in}^{\rm disk}>1$, the radiation fields of NGC~628 have a slower decrease than that of the MW. Overall the MW has a more compact radial distribution than NGC~628.

Instead, the spatial distribution of sSFR is very similar, with both NGC~628 and the MW having a rather constant sSFR beyond $R_{\rm in}^{\rm disk}$, and a supressed sSFR in the centre. Both galaxies have a decreasing sSFR from $R_{\rm in}^{\rm disk}$ to the centre, with the difference that the decrease is monotonic for the Milky Way, and oscillatory for NGC~628. Likewise, the spatial distribution of $F_{\rm young}^{\rm dust}$ is also constant beyond the inner radius, for both NGC~628 and the MW. Both galaxies have the dust heating in the centre dominated by the old stellar populations from the bulge.

We conclude that, although NGC~628 and the Milky Way have different morphology and show some small differences in some respects, they share many common properties, both for the global emission and for the spatial distributions, with some parameters being even identical. We therefore consider NGC~628 a good analogue of the Milky Way.

\section{Summary and Conclusions}
\label{sec:summary}

We applied the radiative transfer model of Popescu et al. (2011) and the methodology from Thirlwall et al. (2020) to account for the panchromatic azimuthally averaged surface brightness distributions of the face-on spiral NGC~628. We used all available imaging data from GALEX, SDSS, {\it Spitzer} and {\it Hershel} to produce UV/optical/NIR/FIR/submm radial profiles, which we fitted with corresponding model profiles. We obtained an excellent fit to the data, both in terms of the overall amplitude and of the detailed shape (slope) of the profiles. In particular the profiles in the 70-350\,${\mu}$m wavelength range, which are not fitted but predicted, show a very good agreement between model and observations. Even the decrease of the scale-length of the infrared emissivity between the 500 and the 70\,${\mu}$m data is exactly predicted by the model. This is probably the best ever fit to the data obtained with a radiative transfer method. The solution obtained for the distribution of stars, dust and radiation fields provided insights into the intrinsic parameters of this galaxy, which, in turn, allowed us to draw conclusions regarding the evolutionary path of this spiral galaxy. The main results are as follows:

\begin{itemize}
\item
The galaxy is marginally optically thick in the centre, with $\tau^f_B(0)=1.16$. The derived dust-to-gas ratio is $D/G=0.0069\pm 0.0005$.

\item
The dust heating is mainly powered by the young stellar population. The fraction of stellar light coming from young stars 
that heats the dust,  $F_{\rm young}^{\rm dust}$, is at a rather constant value of $(0.75-0.8)$ for radii beyond the inner radius of the disk.
In the very centre the heating of the dust is dominated by the old stellar populations from the bulge, with 
$F_{\rm old}^{\rm dust}=0.65$. The global value is $F_{\rm young}^{\rm dust}=0.71$.

\item
The radial profile of SFR surface density, $\Sigma_{\rm SFR}$, shows a pronounced peak at the position of the inner thin disk. Beyond the inner radius of the stellar disk, $\Sigma_{\rm SFR}$ decreases exponentially, with a scale-length $h_{ \Sigma_{\rm SFR}}=(2.8\pm 0.2)$\,kpc. The average value is ${\rm \Sigma_{\rm SFR}}=(1.6\pm0.2)\times10^{-3}\,{\rm M}_{\odot}\,{\rm yr}^{-1}$\,kpc$^{-2}$ within 20\,kpc radius. The total star-formation rate of NGC~628 is ${\rm SFR}=(2.0\pm0.15)\,{\rm M}_{\odot}{\rm yr}^{-1}$.

\item
The radial profile of the stellar mass surface density, $\Sigma_{\rm M*}$, has a pronounced peak in the centre,
followed by an exponential decline with a scale-length $h_{\Sigma_{\rm M*}}=2.7\pm0.2$\,kpc.

\item The radial profile of specific SFR, sSFR, has an almost constant values between the inner radius of the disk and the outer truncation radius, and a non-monotonic decrease towards the centre.
\item The radial profile of dust mass surface density, $\Sigma_{\rm M_d}$, has a peak at the position of the inner thin disk, and an exponential decrease at larger radii, with a scale-length of $h_{\Sigma_{\rm M_{\rm d}}}=7.1\pm 0.5$\,kpc. 
\item We interpret our finding regarding the spatial distributions  of SFR, $M_*$ and $M_{\rm dust}$ in terms of NGC~628 evolving without an efficient inside-out mechanism for disk growth. Instead, efficient mechanisms for transporting dust grains to the outer disk must operate, either from the inner disk or from an external source.
\item
The slope of the global UV attenuation curve of NGC~628 was found to be steeper than that of the inner disk, but shallower than that of the outer disk. This trend was explained in terms of opacity effects.

\item
We quantify projection effects on the surface brightness photometry of NGC~628, by comparing results from the GALFIT bulge-disk decomposition and our RT results. We find the projection effects to diminish the measured Sersic index of the bulge by 0.6 (with respect to the intrinsic index).
\item
In the NIR we find a small level of negative attenuation due to scattering. 
\item
Despite the different morphology, we identify many of the intrinsic properties of NGC~628 to be remarkable similar or identical to those of the Milky Way. Because of this we consider NGC~628 to be, in many respects, an analogue of the Milky Way.
\end{itemize}

The success of our axi-symmetric RT codes and methodology to self-consistently fit both the large-scale geometry and  spectral luminosity of  M33 (Thirlwall et al. 2020) and NGC~628 (this paper) allows us to conclude that these models are powerful tools for decodding face-on spiral galaxies. There is still a need to explore how far from the well behaved spiral morphology a galaxy needs to depart (how irregular a galaxy needs to be) before these models start to reach their limitations. This will be explored in future work. 

\section*{Acknowledgements}

We would like to thank an anonymous referee for very useful suggestions that improved the quality of the paper.
MR acknowledges critical discussions with Dr. Richard Tuffs on the topics presented in this paper, during his scientific visits at the Max Planck Institut f\"ur Kernphysik. CI acknowledges support from a  STFC studentship grant.  
This work is based in part on observations made with the NASA Galaxy Evolution Explorer. GALEX is operated for NASA by the California Institute of Technology under NASA contract NAS5-98034.  
This research has made use of the NASA/IPAC Infrared Science Archive, which is operated by the Jet Propulsion Laboratory, California Institute of Technology, under contract with the National Aeronautics and Space Administration. 
This work has also made use of data products from the Two Micron All Sky Survey, which is a joint project of the University of Massachusetts and the Infrared Processing and Analysis Center/California Institute of Technology, funded by the National Aeronautics and Space Administration and the National Science Foundation. 
This work is based in part on observations made with the {\it Spitzer} Space Telescope, which is operated by the Jet Propulsion Laboratory, California Institute of Technology under a contract with NASA.
We also utilise observations performed with the ESA {\it Herschel} Space Observatory (Pilbratt et al. 2010), in particular 
to do photometry using the PACS  (Poglitsch et al. 2010) and SPIRE (Griffin et al. 2010) instruments.

\section*{Data Availability}

The data underlying this article will be shared on reasonable request to the corresponding author.
 








\appendix
\section{The spectral energy distribution and luminosity of the model components of NGC~628}

\begin{table}
\caption{Intrinsic (dust de-attenuated) spectral luminosity densities in W/Hz of the different stellar components of our model.}
  \begin{tabular}{ l c c c c}
    \hline
 $\lambda$ ($\mu$m ) & $L_{\nu}^{\rm bulge}$      & $L_{\nu}^{\rm disk}$      & $L_{\nu}^{\rm tdisk}$     &  $L_{\nu}^{\rm i-tdisk}$ \\
 \hline
0.091                &      -               &       -             & $1.04\times10^{21}$ & $1.01\times10^{19}$\\
0.130                &      -               &       -             & $1.96\times10^{21}$ & $3.02\times10^{19}$\\
0.150                &      -               &       -             & $1.94\times10^{21}$ & $3.07\times10^{19}$\\
0.152                &      -               &       -             & $1.94\times10^{21}$ & $3.09\times10^{19}$\\
0.160                &      -               &       -             & $2.01\times10^{21}$ & $3.25\times10^{19}$\\
0.200                &      -               &       -             & $1.90\times10^{21}$ & $5.90\times10^{19}$\\
0.220                &      -               &       -             & $2.00\times10^{21}$ & $6.48\times10^{19}$\\
0.227                &      -               &       -             & $2.03\times10^{21}$ & $6.61\times10^{19}$\\
0.250                &      -               &       -             & $1.93\times10^{21}$ & $6.45\times10^{19}$\\
0.280                &      -               &       -             & $2.15\times10^{21}$ & $7.48\times10^{19}$\\
0.355                & $4.60\times10^{19}$  & $9.32\times10^{20}$ & $1.95\times10^{21}$ & $5.59\times10^{19}$\\
0.365                & $5.93\times10^{19}$  & $1.20\times10^{21}$ & $2.09\times10^{21}$ & $6.05\times10^{19}$\\
0.440                & $1.05\times10^{20}$  & $3.20\times10^{21}$ & $2.48\times10^{21}$ & $1.22\times10^{20}$\\ 
0.477                & $1.37\times10^{20}$  & $4.07\times10^{21}$ & $2.96\times10^{21}$ & $1.47\times10^{20}$\\
0.564                & $3.05\times10^{20}$  & $5.35\times10^{21}$ & $3.06\times10^{21}$ & $1.99\times10^{20}$\\ 
0.623                & $4.02\times10^{20}$  & $5.79\times10^{21}$ & $3.73\times10^{21}$ & $2.23\times10^{20}$\\
0.762                & $4.74\times10^{20}$  & $7.98\times10^{21}$ & $4.77\times10^{21}$ & $3.09\times10^{20}$\\
0.809                & $5.47\times10^{20}$  & $8.85\times10^{21}$ & $5.03\times10^{21}$ & $3.25\times10^{20}$\\ 
0.913                & $6.27\times10^{20}$  & $1.03\times10^{22}$ & $4.06\times10^{21}$ & $2.62\times10^{20}$\\
1.259                & $9.96\times10^{20}$  & $1.34\times10^{22}$ & $1.65\times10^{21}$ & $5.31\times10^{19}$\\
1.662                & $1.16\times10^{21}$  & $1.50\times10^{22}$ & $1.50\times10^{21}$ & $4.82\times10^{19}$\\
2.200                & $7.96\times10^{21}$  & $1.18\times10^{22}$ & $1.29\times10^{21}$ & $4.16\times10^{19}$\\
3.368                & $6.80\times10^{20}$  & $5.28\times10^{21}$ & $4.48\times10^{20}$ & $1.78\times10^{19}$\\
3.550                & $6.12\times10^{20}$  & $5.43\times10^{21}$ & $4.02\times10^{20}$ & $1.60\times10^{19}$\\
4.493                & $4.15\times10^{20}$  & $2.82\times10^{21}$ & $3.15\times10^{20}$ & $1.00\times10^{19}$\\
4.618                & $3.79\times10^{20}$  & $2.81\times10^{21}$ & $2.98\times10^{20}$ & $9.44\times10^{18}$\\
5.731                & $2.05\times10^{20}$  & $2.60\times10^{21}$ & $1.95\times10^{20}$ & $6.12\times10^{18}$\\
 \hline\\
 \end{tabular}
 \label{tab:intrlum}
 \end{table}


\begin{table}
\caption{Intrinsic (dust de-attenuated) stellar luminosities in [W] of the model components.}
  \begin{tabular}{ll}
    \hline
$L_{\rm s}^{\rm total}$ & $12.8\times 10^{36}$\\
$L_{\rm s}^{\rm total, uv}$ & $5.14\times 10^{36}$\\
$L_{\rm s}^{\rm disk}$  & $5.54\times 10^{36}$
\\
$L_{\rm s}^{\rm bulge}$ & $0.37\times 10^{36}$
\\
$L_{\rm s}^{\rm tdisk}$ & $6.63\times 10^{36}$\\
$L_{\rm s}^{\rm i-tdisk}$ & $0.21\times 10^{36}$ \\
\hline
 \end{tabular} 
 \label{tab:lumin_stellar}
 \end{table}

\begin{table}
\caption{Dust luminosities in [W] of the model components.}
  \begin{tabular}{ll}
    \hline
$L_{\rm d}^{\rm total}$ & $5.03\times 10^{36}$ \\
$L_{\rm d}^{\rm diff}$ & $3.97\times 10^{36}$ \\
$L_{\rm d}^{\rm clumpy}$ & $1.05\times 10^{36}$ \\
\hline
 \end{tabular}
 \label{tab:lumin_dust}
 \end{table}

\bsp	


\section{Bulge-Disk decomposition with GALFIT and RT model}

  \begin{figure*}
\includegraphics[width=0.7\textwidth,angle=-0]{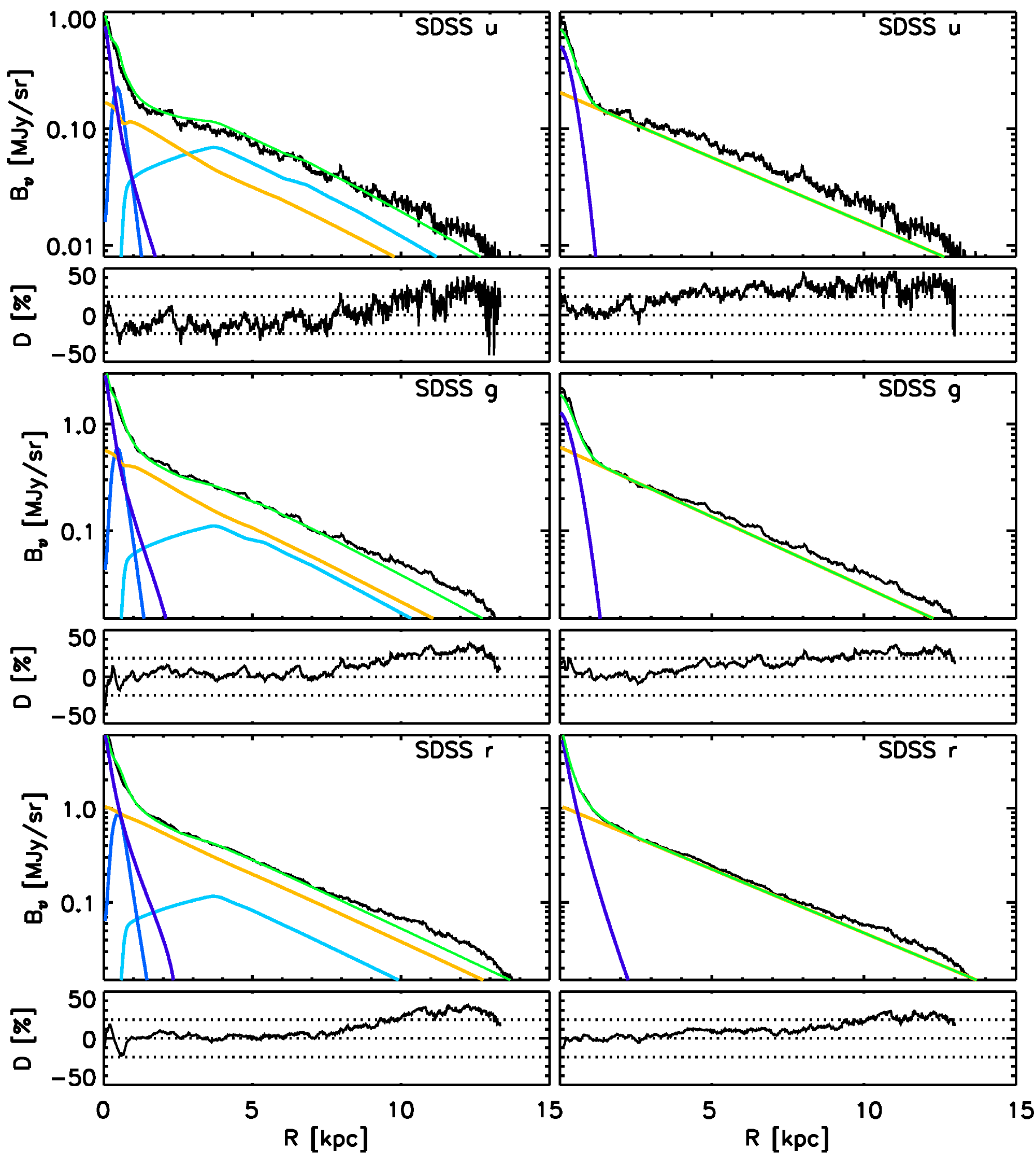}
\caption{Comparison between the model surface brightness profiles obtained from our RT analysis using multicomponent fits (left) and the corresponding ones obtained from GALFIT using B/D decomposition (right). The comparison is done at the wavelengths corresponding to the SDSS ugr bands.}
\label{fig:galfit_SDSS}
\end{figure*}

 \begin{figure*}
\includegraphics[width=0.7\textwidth,angle=-0]{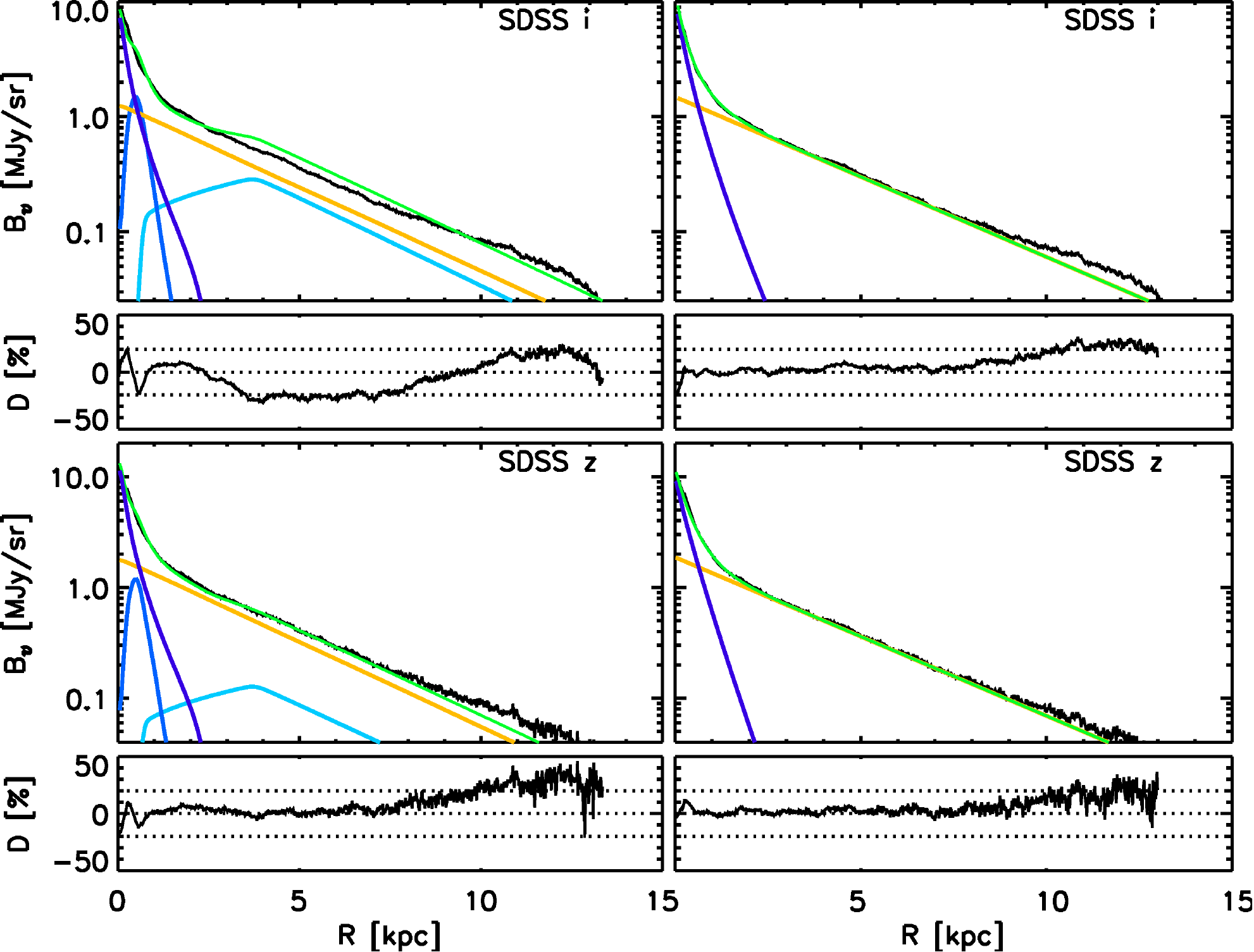}
\caption{Same as in Fig.~\ref{fig:galfit_SDSS}, but for the i and z bands.}.
\label{fig:galfit_SDSSi}
\end{figure*}

\begin{figure*}
\includegraphics[width=0.7\textwidth,angle=-0]{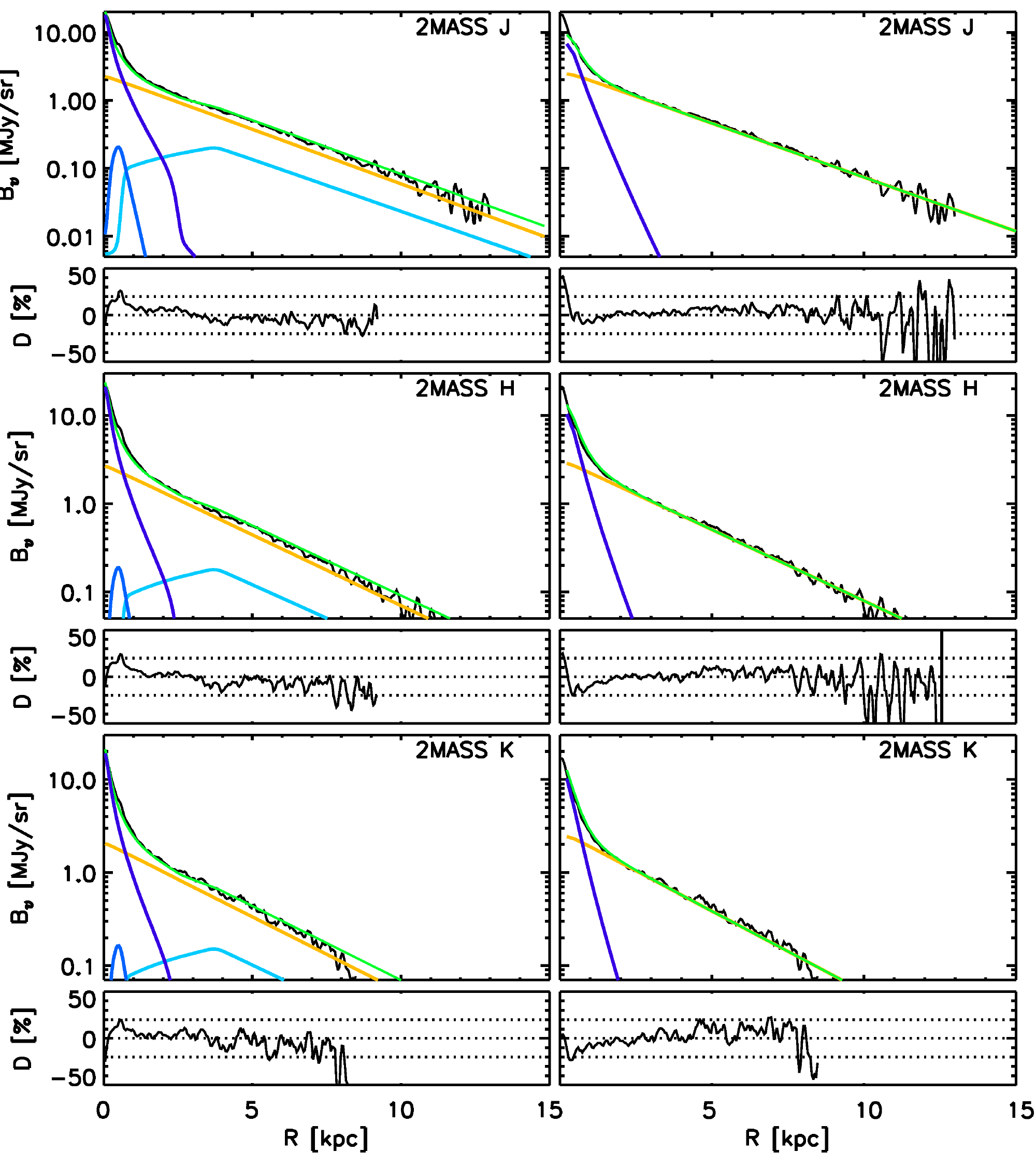}
\caption{Same as in Fig.~\ref{fig:galfit_SDSS}, but for the JHK bands.}
\label{fig:galfit_2MASS}
\end{figure*}

 \label{lastpage}
\end{document}